\documentclass[12pt]{iopart}
\usepackage{graphicx}
\usepackage{amssymb}
\begin{document}
\begin{flushright}
{\small
ITP-UU-04/40\\
SPIN-04/23\\
gr-qc/0411093\\
}
\end{flushright}
\title{Categorizing Different
Approaches to the Cosmological Constant Problem}
\author{Stefan Nobbenhuis}
\address{\ Institute for Theoretical Physics \\
Utrecht University, Leuvenlaan 4\\ 3584 CC Utrecht, the
Netherlands\medskip \\ and
\medskip \\ Spinoza Institute \\ Postbox 80.195 \\ 3508 TD
Utrecht, the Netherlands \smallskip}
\ead{S.J.B.Nobbenhuis@phys.uu.nl}
\begin{abstract}
We have found that proposals addressing the old cosmological
constant problem come in various categories. The aim of this paper
is to identify as many different, credible mechanisms as possible
and to provide them with a code for future reference. We find that
they all can be classified into five different schemes of which we
indicate the advantages and drawbacks.
\end{abstract}

\maketitle
\newpage
\tableofcontents  \pagestyle{headings}

\section{Statement of the Problem}

It is clear that the cosmological constant problem is one of the
major obstacles to further progress for both particle physics and
cosmology. Actually, after the remarkable discoveries and
subsequent confirmations starting in 1997
(SN)\cite{Riess1998,Garnavich1998,Filippenko1998,Perlmutter1998,Perlmutter1998-2,Riess2000,Riess2001,Tonry2003,Knop2003,Barris2003,Riess2004}
(WMAP) \cite{Bennett2003,Spergel2003} (Boomerang)
\cite{Netterfield2001,deBernardis2000} (SdSS)
\cite{Tegmark2003,Afshordi2003} (Hubble) \cite{Freedman2000} that
the universe really is accelerating its expansion, there appear to
be at present at least three cosmological constant problems. In a
nutshell these are: Why is the cosmological constant so small, why
is it then not exactly equal to zero and why is its energy density
today of the same order of magnitude as the matter energy density?
Although the recent observations concerning the accelerated
expansion are usually attributed to a small, non-vanishing
$\Lambda$, alternatives have been suggested, some of which we
briefly discuss.

In this overview however, we will be mainly concerned with the
first of these questions, the so-called ``old cosmological
constant problem''. To phrase it more precisely, the question is
why is the effective cosmological constant, $\Lambda_{eff}$,
defined as $\Lambda_{eff} = \Lambda + 8\pi G\langle\rho\rangle$ so
close to zero\footnote{Note that using this definition we use
units in which the cosmological constant has dimension
$\mbox{GeV}^2$ throughout. Our metric convention is $(-+++)$.}.
Or, in other words, why is the vacuum state of our universe (at
present) so close to the classical vacuum state of zero energy, or
perhaps better, why is the resulting four-dimensional curvature so
small, or why does Nature prefer a flat spacetime?

The different contributions to the vacuum energy density coming
from ordinary particle physics and graviton loops, would naively
give a value for $\langle\rho\rangle$ of order $M_P^4$ (assuming a
Planck-scale cutoff for the standard model), which then would have
to be (nearly) cancelled by the unknown `bare' value of $\Lambda$.
Note at this point that only the $\Lambda_{eff}$ is observable,
not $\Lambda$.

This cancellation has to be better than about 120~decimal places
if we compare the zero-point energy of a scalar field, using the
Planck scale as a cut-off, and the experimental value of
$\rho_{vac}=\langle\rho\rangle + \Lambda/8\pi G$, being $10^{-47}
\mbox{GeV}^4$. As is well known, even if we take a TeV scale
cut-off the difference between experimental and theoretical
results still requires a fine-tuning of about 59 orders of
magnitude. This magnificent fine-tuning seems to suggest that we
miss an important point here. In this paper we give an overview of
the main ideas that have appeared in trying to figure out what
this point might be.

We have found that proposals addressing this problem come in
various categories. The aim of this paper is to identify as many
different, credible mechanisms as possible and to provide them
with a code for future reference. Our identification code will
look as follows, see table (\ref{idcode}).\newpage
\begin{table}[h!]
\caption{Classification of different approaches. Each of them can
also be thought of as occurring 1) Beyond 4D, or 2) Beyond Quantum
Mechanics, or both.}\label{idcode}
\begin{center}
\begin{tabular}{|l|l|} \hline
Type 0: Just Finetuning &  \\
\hline\hline
Type I: Symmetry; A: Continuous & a) Supersymmetry \\
\hline & b) Scale invariance \\
\hline & c) Conformal Symmetry \\
\hline  $\qquad\qquad\qquad\quad$B: Discrete & d) Imaginary Space \\
\hline & e) Energy $\rightarrow$ -Energy \\
\hline & f) Holography\\
\hline & g) Sub-super-Planckian \\
\hline & h) Antipodal Symmetry \\
\hline & i) Duality Transformations \\
\hline\hline
Type II: Back-reaction Mechanism & a) Scalar \\
\hline
& b) Gravitons\\
\hline
& c) Screening Caused by Trace Anomaly  \\
\hline
& d) Running CC from Renormalization Group  \\
\hline\hline Type III: Violating Equiv. Principle & a) Non-local Gravity, Massive Gravitons\\
\hline
& b) Ghost Condensation \\
\hline
& c) Fat Gravitons \\
\hline & d) Composite graviton as Goldst. boson \\
\hline\hline Type IV: Statistical Approaches& a) Hawking Statistics \\
\hline
& b) Wormholes\\
\hline
& c) Anthropic Principle, Cont.\\
\hline & d) Anthropic Principle, Discrete \\ \hline
\end{tabular}
\end{center}
\end{table}
In other words, an approach examining 6-dimensional supersymmetry
for a  solution will be coded Type IAa1.

For reviews on the history of the cosmological constant (problem)
and many phenomenological considerations, see
\cite{Dolgov:1989vb,Weinbergreview,Sahni1999,Carroll2000,Weinberg2000,Padmanabhan2002,Peebles2002,Straumann2002,Ellwanger2002,Yokoyama2003}.

\section{Type 0: Finetuning}

One can set the cosmological constant to any value one likes, by
simply adjusting by hand the value of the bare cosmological
constant to all, classical and quantum mechanical, contributions
to the vacuum energy. No further explanation then is needed. This
fine-tuning has to be precise to better than at least 59~decimal
places (assuming some TeV scale cut-off), but that is of course
not a practical problem. Since we feel some important aspects of
gravity are still lacking in our understanding and nothing can be
learned from this `mechanism', we do not consider this to be a
physical solution. However, it is a possibility that we can not
totally ignore and it is mentioned here just for sake of
completeness.

\section{Type I: Symmetry Principle}

A natural way to understand the smallness of a physical parameter
is in terms of a symmetry that altogether forbids any such term to
appear. This is also often referred to as `naturalness': a theory
obeys naturalness only if all of its small parameters would lead
to an enhancement of its exact symmetry group when replaced by
zero. Nature has provided us with several examples of this. Often
mentioned in this respect is the example of the mass of the
photon. The upper bound on the mass (squared) of the photon from
terrestrial measurements of the magnetic field yields:
\begin{equation}
m_{\gamma}^2\lesssim\mathcal{O}(10^{-50})\mbox{GeV}^2.
\end{equation}
The most stringent estimates on $\Lambda_{eff}$ nowadays give:
\begin{equation}
\Lambda_{eff}\lesssim\mathcal{O}(10^{-84})\mbox{GeV}^2
\end{equation}
We `know' the mass of the photon to be in principle exactly equal
to $0$, because due to the $U(1)$ gauge symmetry of QED, the
photon has only two physical degrees of freedom (helicities). In
combination with Lorentz invariance this sets the mass equal to
zero. A photon with only two transverse degrees of freedom can
only get a mass if Lorentz invariance is broken. This suggests
that there might also be a symmetry acting to keep the effective
cosmological constant an extra~34 orders of magnitude smaller.

A perhaps better example to understand the smallness of a mass is
chiral symmetry. If chiral symmetry were an exact invariance of
Nature, quark masses and in particular masses for the pseudoscalar
mesons $(\pi,K,\eta)$ would be zero. The spontaneous breakdown of
chiral symmetry would imply pseudoscalar Goldstone bosons, which
would be massless in the limit of zero quark mass. The octet
$(\pi,K,\eta)$ would be the obvious candidate and indeed the pion
is by far the lightest of the mesons. Making this identification
of the pion as a pseudo-Goldstone boson associated with
spontaneous breaking of chiral symmetry, we can understand why the
pion-mass is so much smaller than for example the proton mass.

\subsection{Supersymmetry}

One symmetry with this desirable feature is supersymmetry. The
quantum corrections to the vacuum coming from bosons are of the
same magnitude, but opposite sign compared to fermionic
corrections, and therefore cancel each other. The vacuum state in
an exactly supersymmetric theory has zero energy. However,
supersymmetric partners of the Standard Model particles have not
been found, so standard lore dictates that SUSY is broken at least
at the TeV scale, which induces a large vacuum energy.

One often encounters some numerology in these scenarios, e.g.
\cite{Banks2000little}, linking the scale of supersymmetry
breaking $M_{susy}$, assumed to be of order~TeV, and the Planck
mass $M_P$, to the cosmological constant. Experiment indicates:
\begin{equation}
M_{susy}\sim
M_P\left(\frac{\Lambda}{M_P^2}\right)^{\alpha},\quad\quad\mbox{with}\quad\alpha=\frac{1}{8}
\end{equation}
The standard theoretical result however indicates
$M_{susy}\sim\Lambda ^{1/2}$.

However, to discuss the cosmological constant problem, we need to
bring gravity into the picture. This implies making the
supersymmetry transformations local, leading to the theory of
supergravity or SUGRA for short, where the situation is quite
different. In exact SUGRA the lowest energy state of the theory,
generically has negative energy density: the vacuum of
supergravity is AdS\footnote{This negative energy density can also
be forbidden by postulating an unbroken R-symmetry. $D=11$ SUGRA
is a special case; its symmetries implicitly forbid a CC term, see
\cite{DeserBautier1997}.}. This has inspired many to consider
so-called no-scale supergravity models. See \cite{Weinbergreview}
or supersymmetry textbooks such as \cite{BailinLove} for excellent
reviews.

The important point is that there is an elegant way of
guaranteeing a flat potential, with $V=0$ after susy-breaking, by
using a nontrivial form of the K\"{a}hler potential $G$. For a
single scalar field $z$ we have:
\begin{eqnarray}
V &=&
e^{G}\left[\frac{\partial_{z}G\partial_{z^{\ast}}G}{\partial_{z}\partial_{z^{\ast}}G}
- 3\right]\nonumber\\
&=&
\frac{9e^{4G/3}}{\partial_{z}\partial_{z^{\ast}}G}(\partial_{z}\partial_{z^{\ast}}e^{-G/3}),
\end{eqnarray}
where $\kappa^2$, the gravitational constant, has been set equal
to one. A flat potential with $V=0$ is obtained if the expression
in brackets vanishes for all $z$, which happens if:
\begin{equation}
G = -3\log(z + z^{\ast}),
\end{equation}
and one obtains a gravitino mass:
\begin{equation}
m_{3/2} = \langle e^{G/2}\rangle = \langle (z +
z^{\ast})^{-3/2}\rangle,
\end{equation}
which as required is not fixed by the minimization of $V$. Thus
provided we are prepared to choose a suitable, nontrivial form for
the K\"{a}hler potential $G$, it is possible to obtain zero CC.
Moreover, the gravitino mass is left undetermined; it is fixed
dynamically through non-gravitational radiative corrections. The
minimum of the effective potential occurs at:
\begin{equation}
V_{eff}\approx -(m_{3/2})^{4},
\end{equation}
where in this case after including the observable sector and soft
symmetry-breaking terms we will have $m_{3/2}\approx M_{W}$. Such
a mass is ruled out cosmologically \cite{Grifols} and so other
models with the same ideas have been constructed that allow a very
small mass for the gravitino, also by choosing a specific
K\"{a}hler potential, see \cite{Lahanas:1986uc}.

That these constructions are possible is quite interesting and in
the past there has been some excitement when superstring theory
seemed to implicate precisely the kinds of K\"{a}hler potential as
needed here, see for example \cite{Witten1985}. However, that is
not enough, these simple structures are not expected to hold
beyond zeroth order in perturbation theory.

\subsubsection{Unbroken SUSY}

To paraphrase Witten \cite{EWUnbrokensusy}: ``Within the known
structure of physics, supergravity in four dimensions leads to a
dichotomy: either the symmetry is unbroken and bosons and fermions
are degenerate, or the symmetry is broken and the vanishing of the
CC is difficult to understand''. However, as he also argues in the
same article, in $2+1$ dimensions, this unsatisfactory dichotomy
does not arise: SUSY can explain the vanishing of the CC without
leading to equality of boson and fermion masses, see also
\cite{EWCCfromString}.

The argument here is that in order to have equal masses for the
bosons and fermions in the same supermultiplet one has to have
unbroken global supercharges. These are determined by spinor
fields which are covariantly constant at infinity. The
supercurrents $J^{\mu}$ from which the supercharges are derived
are generically not conserved in the usual sense, but covariantly
conserved: $D_{\mu}J^{\mu} = 0$. However, in the presence of a
covariantly constant spinor ($D_{\mu}\epsilon =0$), the conserved
current $\bar{\epsilon}J^{\mu}$ can be constructed and therefore,
a globally conserved supercharge:
\begin{equation}
Q = \int d^3 x\bar{\epsilon}J^0.
\end{equation}
But in a $2+1$ dimensional spacetime any state of non-zero energy
produces a geometry that is asymptotically conical at infinity
\cite{DeserJackiwtHooft}. The spinor fields are then no longer
covariantly constant at infinity \cite{Henneaux1984} and so even
when supersymmetry applies to the vacuum and ensures the vanishing
of the vacuum energy, it does not apply to the excited states.
This is special to $2+1$-dimensions. Explicit examples have been
constructed in
\cite{BeckerBeckerStrominger,Dvalifbdegen,Edelstein95,Edelstein96}.
Two further ideas in this direction, one in $D<4$ and one in $D>4$
are \cite{GregoryRubakovSibiryakov,CsakiErlichHollowood}, however
the latter later turned out to be internally inconsistent
\cite{DvaliGababadzePorrati}.

In any case, what is very important is to make the statement of
`breaking of supersymmetry' more precise. As is clear, we do not
observe mass degeneracies between fermions and bosons, therefore
supersymmetry, even if it were a good symmetry at high energies
between excited states, is broken at lower energies. However, and
this is the point, as the example of Witten shows, the issue of
whether we do or do not live in a supersymmetric vacuum state is
another question. In some scenarios it is possible to have a
supersymmetric vacuum state, without supersymmetric excited
states. This really seems to be what we are looking for. The
observations of a small or even zero CC could point in the
direction of a (nearly) supersymmetric vacuum state.

Obviously the question remains how this scenario and the absence
nevertheless of a supersymmetric spectrum can be incorporated in 4
dimensions, where generically spacetime is asymptotically flat
around matter sources, instead of asymptotically conical.

\subsection{Imaginary Space}\label{imaginaryspace}

So far, the most obvious candidate-symmetry to enforce zero vacuum
energy density, supersymmetry, does not seem to work; we need
something else. What other symmetry could forbid a cosmological
constant term? Einstein's equations are:
\begin{equation}
R_{\mu\nu} - \frac{1}{2}g_{\mu\nu}R - \Lambda g_{\mu\nu} = -8\pi
GT_{\mu\nu}
\end{equation}
As was first observed by 't Hooft (unpublished), we can forbid the
cosmological constant term by postulating that the
transformations:
\begin{equation}\label{thooftsym}
\mathbf{x}\rightarrow i\mathbf{x},\quad\quad t\rightarrow
it,\quad\quad g_{\mu\nu}\rightarrow g_{\mu\nu}
\end{equation}
are symmetry operations \footnote{A related suggestion was made in
\cite{Erdem2004}.}. The different objects in Einstein's equations
transform under this as follows:
\begin{eqnarray}
\Gamma^{\lambda}_{\mu\nu} &=& \frac{1}{2}g^{\lambda\rho}
\left[\partial_{\mu}g_{\nu\rho} + \partial_{\nu}g_{\mu\sigma} -
\partial_{\rho}g_{\mu\sigma}\right]
\rightarrow -i\Gamma^{\lambda}_{\mu\nu}\nonumber\\
R_{\mu\nu} &=& \partial_{\nu}\Gamma^{\lambda}_{\mu\lambda} -
\partial_{\lambda}\Gamma^{\lambda}_{\mu\nu} -
\Gamma^{\rho}_{\mu\nu}\Gamma^{\sigma}_{\rho\sigma} +
\Gamma^{\rho}_{\mu\sigma}\Gamma^{\sigma}_{\nu\rho}
\rightarrow -R_{\mu\nu}\nonumber\\
R &=& g^{\mu\nu}R_{\mu\nu} \rightarrow -R\nonumber
\end{eqnarray}
Furthermore we have:
\begin{equation}
T_{\mu\nu}\rightarrow - T_{\mu\nu}
\end{equation}
as long as there are no vacuum terms in the expression for
$T_{\mu\nu}$. So Einstein's equations transforms as:
\begin{equation}
G_{\mu\nu} - \Lambda g_{\mu\nu} = -8\pi G T_{\mu\nu} \rightarrow
-G_{\mu\nu} - \Lambda g_{\mu\nu}= +8\pi GT_{\mu\nu}
\end{equation}
Therefore, if we postulate (\ref{thooftsym}) as a symmetry of
nature, a CC term is forbidden! Classically $E\sim p^2$ and
therefore $E\rightarrow -E$.

However, at first sight, this symmetry does not seem to ameliorate
the situation much, since this transformation is not a symmetry of
the Standard Model. In particular, we have:
\begin{equation}
p^2 = m^2,\quad\quad\mbox{with}\quad p^{\mu} =
i\partial^{\mu}\rightarrow -ip^{\mu}
\end{equation}
Therefore, imposing (\ref{thooftsym}) as a symmetry of nature,
seems to imply that either there exists a copy of all known matter
particles with negative mass squared, or that all particles should
be massless. In the second case, if we take this symmetry
seriously, we should conclude that the smallness of the
cosmological constant and the smallness of particle masses
(relative to the Planck-scale) although of quite a different order
of magnitude, have a common origin.

Another approach is to view this symmetry in combination with
boundary conditions. Generally in quantum field theory we Fourier
transform our field and impose (often periodic) boundary
conditions only on its components in real space. Perhaps the
vacuum state does not have to satisfy boundary conditions. In that
case it would not matter whether one would impose boundary
conditions in either real or imaginary space. Excited states, have
to obey boundary conditions, and would violate the symmetry.

Besides, since the transformation (\ref{thooftsym}) effectively
changes spacelike dimensions into timelike dimensions and vice
versa, a natural playground to study its implications could be a
$2+2$- or $3+3$-dimensional spacetime. The possibility of extra
timelike dimensions is not very often considered, because it is
assumed that the occurrence of tachyonic modes prevents the
construction of physically viable models. However, it was shown in
\cite{Dvaliextratime} that these constraints might not be as
severe as to rule out this option beforehand. Extra timelike
dimensions have been tried before to argue for a vanishing
cosmological constant, see \cite{Berezhiani:2001bn}.

\subsection{Energy $\rightarrow -$ Energy}\label{Energytominus}

Another approach in which negative energy states are considered
has been recently proposed in \cite{KaplanSundrum2005}. Here the
discrete symmetry $E\rightarrow -E$ is imposed explicitly on the
matter fields by adding to the Lagrangian an identical copy of the
normal matter fields, but with an overall minus sign:
\begin{equation}
\mathcal{L} =\sqrt{-g}\left( M_{Pl}^2R - \Lambda_0 +
\mathcal{L}_{matt}(\psi,D_{\mu}) -
\mathcal{L}_{matt}(\hat{\psi},D_{\mu}) +\dots\right),
\end{equation}
where $\Lambda_0$ is the bare cosmological constant. The
Lagrangian with fields $\hat{\psi}$ occurring with the wrong sign
is referred to as the ghost sector. The two matter sectors have
equal but opposite vacuum energies, and therefore cancelling
contributions to the cosmological constant.

Crucial in this reasoning is that there is no coupling other than
gravitational between the normal matter fields and their ghost
counterparts, otherwise the Minkowski vacuum would not be stable.
This gravitational coupling moreover has to be sufficiently small
in order to suppress the gravitationally induced interactions
between the two sectors and to make sure that the quantum
gravitational corrections to the bare cosmological constant are
kept very small. It is therefore necessary to impose a UV cutoff
on these contributions of order $10^{-3}$~eV, corresponding to a
length scale of about 100~microns\footnote{In section
(\ref{fatgravitons}) a proposal by one of the authors of
\cite{KaplanSundrum2005} is discussed in which such a cutoff is
argued to arise from the graviton not being a point-like particle
but having this finite size.}.

Moreover, in order to ensure stability of the vacuum, also some
new Lorentz symmetry violating physics is required to suppress
processes where normal matter particles and ghosts emerge from the
vacuum. In addition, one also has to assume that the ghost sector
is rather empty, compared to the normal matter sector, in order
not to spoil standard cosmology with such an exotic type dark
matter.

\subsection{Scale Invariance, e.g. Conformal Symmetry}

The above symmetry might be viewed as a specific example of the
more general framework of conformal symmetry,
$g_{\mu\nu}\rightarrow f(x^{\mu})g_{\mu\nu}$. Massless particles
are symmetric under a bigger group than just the Lorentz group,
namely, the conformal group. This group does not act as symmetries
of Minkowski spacetime, but under a (mathematically useful)
completion, the ``conformal compactification of Minkowski space".
This group is 15-dimensional and corresponds to $SO(2,4)$, or if
fermions are present, the covering group $SU(2,2)$. Conformal
symmetry forbids any term that sets a length scale, so a
cosmological constant is not allowed, and indeed also particle
masses necessarily have to vanish.

General coordinate transformations and scale invariance, i.e.
$g_{\mu\nu}\rightarrow fg_{\mu\nu}$, are incompatible in general
relativity. The $R\sqrt{-g}$ term in the Einstein-Hilbert action
is the only quantity that can be constructed from the metric
tensor and its first and second derivatives only, that is
invariant under general coordinate transformations. But this term
is not even invariant under a global scale transformation
$g_{\mu\nu}\rightarrow fg_{\mu\nu}$ for which $f$ is constant. $R$
transforms with Weyl weight $-1$ and $\sqrt{-g}$ with weight $+2$.
There are two ways to proceed to construct a scale invariant
action: introducing a new scalar field
\cite{Deser1970,Dirac:1973gk}, that transforms with weight $-1$,
giving rise to so-called scalar-tensor theories, or consider
Lagrangians that are quadratic in the curvature scalar. We
consider the second. Pioneering work in this direction was done in
\cite{Zee:1978wi,Zee:1981qn,Zee:1980sj,Zee:1981ff,Zee:1983mj}. See
for example \cite{Booth,Barbour} for some resent studies and many
references.

Gravity can be formulated under this bigger group, leading to
``Conformal gravity", defined in terms of the Weyl tensor, which
corresponds to the traceless part of the Riemann tensor:
\begin{eqnarray}\label{weylaction}
S_G &=& -\alpha\int d^4
x\sqrt{-g}C_{\lambda\mu\nu\kappa}C^{\lambda\mu\nu\kappa}\nonumber\\
&=& -2\alpha \int d^4 x\sqrt{-g}\left(R_{\mu\nu}R^{\mu\nu}
-\frac{1}{3}R^2\right) +(\mbox{boundary}\;\mbox{terms}),
\end{eqnarray}
where $C^{\mu\nu\lambda\kappa}$ is the conformal Weyl tensor, and
$\alpha$ is a dimensionless gravitational coupling constant. Thus
the Lagrangian is quadratic in the curvature scalar and generates
field equations that are fourth-order differential equations. Note
that the a cosmological constant term is not allowed, since it
violates conformal invariance. Based on the successes of gauge
theories with spontaneously broken symmetries and the generation
of the Fermi-constant, one may suggest to also dynamically induce
the Einstein action with its Newtonian constant as a macroscopic
limit of a microscopical conformal theory. This approach has been
studied especially by Mannheim and Kazanas, see
\cite{Mannheim991,Mannheim992,Mannheim993,Mannheim96,Mannheim941,Mannheim942}
to solve the CC problem.

These fourth-order equations reduce to a fourth-order Poisson
equation:
\begin{equation}
\nabla^4B(r) = f(r),
\end{equation}
where $B(r) = -g_{00}(r)$ and the source is given by:
\begin{equation}
f(r) = 3(T^{0}_{\ 0}-T^{r}_{\ r})/4\alpha B(r),
\end{equation}
For a static, spherically symmetric source, conformal symmetry
allows one to put $g_{rr} = -1/g_{00}$ and the exterior solution
to (\ref{weylaction}) can be written \cite{Mannheim942}:
\begin{equation}
g_{rr} = -1/g_{00}=1-\beta(2-3\beta\gamma)/r -3\beta\gamma +\gamma
r -kr^2.
\end{equation}
The non-relativistic potential reads:
\begin{equation}\label{nonrelpotconf}
V(r)=-\beta/r + \gamma r/2
\end{equation}
which for a spherical source can be completely integrated to
yield:
\begin{equation}
B(r>R)=-\frac{r}{2}\int^R_0dr'f(r')r'^2
-\frac{1}{6r}\int^R_0dr'f(r')r'^4.
\end{equation}
Compared to the standard second-order equations:
\begin{equation}
\nabla^2\phi(r)=g(r)\quad\quad\rightarrow\quad\quad\phi(r>R)=-\frac{1}{r}\int^R_0dr'g(r')r'^2
\end{equation}
we see that the fourth-order equations contain the Newtonian
potential in its solution, but in addition also a linear potential
term that one would like to see dominate over Newtonian gravity
only at large distances. The factors $\beta$ and $\gamma$ in for
example (\ref{nonrelpotconf}) are therefore given by:
\begin{equation}
\beta(2-3\beta\gamma) =
\frac{1}{6}\int_0^Rdr'f(r')r'^4\quad\quad,\quad\quad\gamma=-\frac{1}{2}\int_0^Rdr'f(r')r'^2
\end{equation}

Note in passing that in the non-relativistic limit of GR the
$(0,0)$-component, where $R_{ij}\simeq (1/2R-\Lambda)g_{ij}$ and
therefore $R=g^{\mu\nu}R_{\mu\nu}\simeq R_{00}+3(1/2R-\Lambda)$,
or $R\simeq -2R_{00}+6\Lambda$ using also that $R_{00}\simeq
(-1/2)\nabla^2g_{00}$ becomes:
\begin{equation}
\nabla^2\phi = 4\pi G\left(\rho - \frac{\Lambda}{4\pi G}\right),
\end{equation}
the Poisson equation for the normal Newtonian potential modified
with a cosmological constant. This can easily be solved to give:
\begin{equation}
\phi = -\frac{GM}{r} + \frac{1}{6}\Lambda r^2.
\end{equation}

However, modifying gravity only at large distances cannot solve
the cosmological constant problem. The (nearly) vanishing of the
vacuum energy and consequently flat and relatively slowly
expanding spacetime is a puzzle already at distance scales of say
a meter. We could expect deviations of GR at galactic scales,
avoiding the need for dark matter, but at solar system scales GR
in principle works perfectly fine. It seems hard to improve on
this, since the world simply is not scale invariant.

There is also a more serious problem with the scenario of Mannheim
and Kazanas described above. In order for the linear term not to
dominate already at say solar system distances, the coefficient
$\gamma$ has to be chosen very small. Not only does this introduce
a new kind of fine-tuning, it is simply not allowed to chose these
coefficients at will. The linear term will always dominate over
the Newtonian $1/r$-term, in contradiction with the perfect
agreement of GR at these scales. See also \cite{PerlickXu} who
raised the same objection.

This scenario therefore does not work.

\subsubsection{$\Lambda$ as Integration Constant, Unimodular Theory}\label{unimodular}

Another option is to reformulate the action principle in such a
way that a scale dependent quantity like the scalar curvature,
remains undetermined by the field equations themselves. These are
the so-called 'unimodular' theories of gravity, see e.g.
\cite{vanderBijvanDamNg,Unruh}. Note that although the action is
not globally scale invariant, Einstein's equations in the absence
of matter and with vanishing cosmological constant is. The
dynamical equations of pure gravity in other words, are invariant
with respect to global scale transformations, and since we have
that $R=0$, they are scale-free, i.e. they contain no intrinsic
length scale.

There is a way to keep the scale dependence undetermined also
after including matter which also generates a cosmological
constant term. This well-known procedure
\cite{AndersonFinkelstein,vanderBij:1981uw}, assumes:
\begin{equation}
\sqrt{-g} =
\sigma(x)\quad\quad\rightarrow\quad\quad\delta\sqrt{-g}=0,
\end{equation}
where $\sigma(x)$ is a scalar density of weight $+1$. The
resulting field equations are:
\begin{equation}
R_{\mu\nu}-\frac{1}{4}g_{\mu\nu}R = -\kappa\left(T_{\mu\nu} -
\frac{1}{4}g_{\mu\nu}T\right).
\end{equation}
The covariant derivative $D_{\mu}G_{\mu\nu}=D_{\mu}T_{\mu\nu}=0$
still vanishes and from this one obtains:
\begin{equation}
R -\kappa T = -4\Lambda,
\end{equation}
where $\Lambda$ now appears as an integration constant and the
factor of $4$ has been chosen for convenience since substituting
this back we recover the normal Einstein equations with
cosmological constant.

Recently, some arguments have been put forward in which a sort of
unimodular theory is supposed to originate more naturally as a
result of 'the quantum microstructure of spacetime being capable
of readjusting itself, soaking up any vacuum energy', see
\cite{Padmanabhan2004, Padmanabhan2002, Padmanabhan2004b}.

Obviously this does not solve anything, nor does it provide a
better understanding of the cosmological constant. The value of
the integration constant $\Lambda$ has to be inserted by hand in
order to arrive at the correct value.

Besides, sometimes it is concluded that there are two inequivalent
Einstein equations for gravity, describing two theories that are
only equivalent classically, but not quantum mechanically. The
group of canonical transformations is much larger than that of
unitary transformations in Hilbert space, forcing one to quantize
in ``preferred" coordinates. We do not agree with this point of
view. The constraint $g^{\mu\nu}\delta g_{\mu\nu}=0$ just reflects
a choice of coordinates, a certain gauge.

This issue is closely related to the question of the measure of
the quantum gravity functional integral (see discussions by B.S.
DeWitt \cite{DeWitt1967,DeWitt1967b}, 't Hooft \cite{tHooft1978}
and \cite{MazurMottola1989}): Is the integration variable
$g_{\mu\nu}$, $g^{\mu\nu}$ or some other function of the metric?
The differences in the amplitudes for these theories all appear in
the one-loop diagrams, in the form of quartically divergent
momentum-independent ghost loops. These all disappear after
renormalization and therefore the theories are indistinguishable
physically.

\subsection{Holography}

Gravitational holography \cite{tHooft1} limits the number of
states accessible to a system. The entropy of a region generally
grows with its covering area (in Planck units) rather than with
its volume, implying that the dimension of the Hilbert space, i.e.
the number of degrees of freedom describing a region, is finite
and much smaller than expected from quantum field theory.
Considering an infinite contribution to the vacuum energy is not
correct because states are counted that do not exist in a
holographic theory of gravity.

It is a symmetry principle since there is a projection from states
in the bulk-volume, to states on the covering surface.

In \cite{Thomas,CohenKaplanNelson} it is noted that in effective
field theory in a box of size $L$ with UV cutoff $M$ the entropy
$S$ scales extensively, as $S\sim L^3M^3$. A free Weyl fermion on
a lattice of size $L$ and spacing $1/M$ has $4^{(LM)^3}$ states
and entropy\footnote{For bosons the number of states is not
limited by a lattice cutoff alone, so in this argument one has to
limit oneself to fermions. For bosons there are an infinite number
of states, in contradiction to the conjecture of the Holographic
Principle.} $S\sim(LM)^3$. The corresponding entropy density
$s=S/V$ then is $s=M^3$. In $d=4$ dimensions quantum corrections
to the vacuum energy are therefore of order:
\begin{equation}
\rho_{vac} = \frac{\Lambda}{8\pi G} + \langle\rho\rangle =
\frac{\Lambda}{8\pi G} + \mathcal{O}(s^{4/3}),
\end{equation}
since both $\langle\rho\rangle$ and $s$ are dominated by
ultraviolet modes, (see also \cite{Hsu}). Thus finite $s$ implies
finite corrections to $\langle\rho\rangle$.

Using a cutoff $M$, $E\sim M^4L^3$ is the maximum energy for a
system of size $L$. States with $L<R_s\sim E$, or $L>M^{-2}$ (in
Planckian units) have collapsed into a black-hole. If one simply
requires that no state in the Hilbert space exists with $R_s\sim
E> L$, then a relation between the size $L$ of the region,
providing an IR cutoff, and the UV cutoff $M$ can be derived.
Under these conditions entropy grows no faster than $A^{3/4}\sim
L^{3/2}$, with $A$ the area. If these black hole states give no
contribution to $\langle\rho\rangle$, we obtain:
\begin{equation}
\langle\rho\rangle\sim s^{4/3}\sim
\left(\frac{L^{3/2}}{L^3}\right)^{4/3}\sim L^{-2}.
\end{equation}
In \cite{Thomas} this same scaling was obtained by assuming that
$S<A$ as usual, but that the delocalized states have typical
Heisenberg energy $~1/L$:
\begin{equation}
\langle\rho\rangle\sim\frac{s}{L}\sim\frac{L^2}{L^3L}\sim L^{-2}.
\end{equation}
Plugging in for $L$ the observed size of the universe today the
quantum corrections are only of order $10^{-10}~\mbox{eV}^4$.

However, this does not yield the correct equation of state,
\cite{Hsu}. During matter dominated epochs, to which WMAP and
supernova measurements are sensitive, the horizon size grows as
the RW-scale factor, $a(t)^{3/2}$, so the above arguments imply:
\begin{equation}
\Lambda_{eff}(L)\sim a(t)^{-3},
\end{equation}
or, $w\equiv p/\rho =0$ at largest scales, since $\rho(t)\sim
a(t)^{-3(1+w)}$. The data on the other hand give $w< -0.78$~(95\%
CL). In for example \cite{Thomas,CohenKaplanNelson} $\Lambda(L)$
is at all times comparable to the radiation + matter energy
density, which is also argued to give problems for structure
formation \cite{Turner2002}.

Holography-based scenarios thus naively lead to a cosmological
constant that is far less constant than what the data require.
This makes a connection between holography and dark energy a lot
harder to understand\footnote{In \cite{Kelleher2} in a different
context a similar relation between the CC and the volume of the
universe is derived, thus suffering from the same drawbacks.}.

More recently however, another proposal was made \cite{Li2004}
where instead $L$ is taken to be proportional to the size of the
future event horizon:
\begin{equation}
L(t)\sim a(t)\int_t^{\infty}\frac{dt'}{a(t')}
\end{equation}
This $L$ describes the size of the largest portion of the universe
that any observer will see. This could be a reasonable IR cutoff.
It is argued that in this case the equation of state parameter $w$
can be close enough to $-1$ to agree with the data. This relation
is rather ad hoc chosen, and its deeper meaning, if any, still has
to be discovered.

Another reason to discuss holography in the context of the
cosmological constant problem lies in trying to reconcile string
theory with the apparent observation of living in a de Sitter
spacetime. The discussion centers around the semi-classical result
that de Sitter space has a finite entropy, inversely related to
the cosmological constant, see for example \cite{Banks2000-1}.
Thus one may reason that de Sitter space should be described by a
theory with a finite number of independent quantum states and that
a theory of quantum gravity should be constructed with a finite
dimensional Hilbert space. In this reasoning a cosmological
constant should be understood as a direct consequence of the
finite number of states in the Hilbert space describing the world.
Ergo, the larger the cosmological constant, the smaller the
Hilbert space. However, in \cite{BoussoMeyersDewolfe} it is argued
that this relation between the number of degrees of freedom and
the CC is not so straightforward.

\subsection{``Symmetry" between Sub- and Super-Planckian Degrees of
Freedom}

This rather speculative reasoning originates from a comparison
with condensed matter physics and is due to Volovik, see for
example
\cite{Volovik2000,Volovik2001a,Volovik2001b,Volovik2002,Volovik2003a,Volovik2003b,Volovik2004,Volovik2005}.
The vacuum energy of superfluid $^4$Helium, calculated from an
effective theory containing phonons as elementary bosonic
particles and no fermions is:
\begin{equation}
\rho_{\Lambda} = \sqrt{-g}E^{4}_{Debye}
\end{equation}
with $g$ the determinant of the acoustic metric, since $c$ is now
the speed of sound, and $E_{Debye}=\hbar c/a$, with $a$ the
interatomic distance, which plays the role of the Planck length.
However, in the condensed matter case, the full theory exists: a
second quantized Hamiltonian describing a collection of a
macroscopic number of structureless $^4$Helium bosons or
$^3$Helium fermions, in which the chemical potential $\mu$ acts as
a Lagrange multiplier to ensure conservation of the number of
atoms:
\begin{eqnarray}
H-\mu N &=&\int
d\mathbf{x}\psi^{\dag}(\mathbf{x})\left[-\frac{\nabla^2}{2m}-\mu\right]\psi(\mathbf{x})\nonumber\\
&+& \int d\mathbf{x}d\mathbf{y}
V(\mathbf{x}-\mathbf{y})\psi^{\dag}(\mathbf{x})\psi^{\dag}(\mathbf{y})\psi(\mathbf{y})\psi(\mathbf{x}).
\end{eqnarray}

Using this Hamiltonian $H$ to calculate the energy density of the
ground state we get:
\begin{equation}
E_{vac} = E - \mu N = \langle\mbox{vac}|H-\mu N|\mbox{vac}\rangle
\end{equation}
An overall shift of the energy in $H$ is cancelled in a shift of
the chemical potential. Exact calculation shows that not only the
low energy degrees of freedom from the effective theory, the
phonons, but also the higher energy, ``trans-Planckian" degrees of
freedom have to be taken into account.

Besides, for a liquid of $N$ identical particles at temperature
$T$ in a volume $V$ in equilibrium, the relation between the
energy $E$ and pressure $P$ is given by the Gibbs-Duhem equation:
\begin{equation}
E = TS +\mu N - PV.
\end{equation}
Therefore at $T=0$ the energy density of the ground state becomes:
\begin{equation}
\rho_{vac}\equiv \frac{E_{vac}}{V} = -P_{vac},
\end{equation}
the same equation of state as for the vacuum state in GR. Using
just thermodynamic arguments, it is argued that in the infinite
volume, zero temperature limit, this gives exactly zero vacuum
energy density as long as there are no external forces, i.e. no
pressure acting on the quantum liquid. And assuming there is no
matter, no curvature and no boundaries which could give rise to a
Casimir effect \cite{Volovik2001a}.

The conclusion therefore is that, if these thermodynamic arguments
are also valid in a gravitational background for the universe as a
whole and up to extremely high energies, one would expect a
perfect cancellation between sub- and super-Planckian degrees of
freedom contributing to the vacuum energy, resulting in zero
cosmological constant.

Moreover, it is also argued that a non-zero cosmological constant
arises from perturbations of the vacuum at non-zero temperature.
The vacuum energy density would be proportional to the matter
energy density, solving the coincidence problem as well.

A similar result is obtained by \cite{KleinertZaanen}. In their
formulation the world is like a crystal. The atoms of the crystal
are in thermal equilibrium and exhibit therefore zero pressure,
making the cosmological constant equal to zero.

Both approaches strongly depend on the quantum systems reaching
their equilibrium state. However, in the presence of a
cosmological constant, the matter in the universe never reaches
its equilibrium state \cite{ShapiroSola2000}.

\subsection{Interacting Universes, Antipodal Symmetry}

This is an approach developed by Linde \cite{Linde,Linde6} arguing
that the vacuum energy in our universe is so small because there
is a global interaction with another universe where energy
densities are negative. Consider the following action of two
universes with coordinates $x_{\mu}$ and $y_{\alpha}$
respectively, ($x_{\mu}, y_{\alpha} = 0,1,\dots,3$) and metrics
$g_{\mu\nu}(x)$ and $\bar{g}_{\alpha\beta}(y)$, containing fields
$\phi(x)$ and $\bar{\phi}(y)$:
\begin{equation}
S = N\int
d^{4}xd^{4}y\sqrt{g(x)}\sqrt{\bar{g}(y)}\left[\frac{M_{P}^{2}}{16\pi}R(x)
+ L(\phi(x)) - \frac{M_{P}^{2}}{16\pi}R(y) -
L(\bar{\phi}(y))\right],
\end{equation}
and where $N$ is some normalization constant. This action is
invariant under general coordinate transformations in each of the
universes separately. The important symmetry of the action is
$\phi(x)\rightarrow\bar{\phi}(x)$,
$g_{\mu\nu}(x)\rightarrow\bar{g}_{\alpha\beta}(x)$ and under the
subsequent change of the overall sign: $S\rightarrow -S$. He calls
this an antipodal symmetry, since it relates states with positive
and negative energies. As a consequence we have invariance under
the change of values of the effective potentials
$V(\phi)\rightarrow V(\phi) + c$ and $V(\bar{\phi})\rightarrow
V(\bar{\phi}) + c$ where $c$ is some constant. Therefore nothing
in this theory depends on the value of the effective potentials in
their absolute minima $\phi_{0}$ and $\bar{\phi}_{0}$. Note that
because of the antipodal symmetry $\phi_{0} = \bar{\phi}_{0}$ and
$V(\phi_{0}) = V(\bar{\phi}_{0})$.

In order to avoid the troublesome issues of theories with negative
energy states, one has to assume that there can be no interactions
between the fields $\phi(x)$ and $\bar{\phi}(y)$. Therefore also
the equations of motion for both fields are the same and
similarly, also gravitons from both universes do not interact.

However some interaction does occur. The Einstein equations are:
\begin{eqnarray}
R_{\mu\nu}(x) - \frac{1}{2}g_{\mu\nu}R(x) &=& -8\pi GT_{\mu\nu}(x)
- g_{\mu\nu}\langle\frac{1}{2}R(y) + 8\pi
GL(\bar{\phi}(y))\rangle\\
R_{\alpha\beta}(y) - \frac{1}{2}\bar{g}_{\alpha\beta}R(y) &=&
-8\pi GT_{\alpha\beta}(y) -
\bar{g}_{\alpha\beta}\langle\frac{1}{2}R(x) + 8\pi
GL(\phi(x))\rangle.
\end{eqnarray}
Here $T_{\mu\nu}$ is the energy-momentum tensor of the fields
$\phi(x)$ and $T_{\alpha\beta}$ the energy-momentum tensor for the
fields $\bar{\phi}(y)$ and the averaging means:
\begin{eqnarray}
\langle R(x)\rangle &=& \frac{\int d^{4}x\sqrt{g(x)}R(x)}{\int
d^{4}x\sqrt{g(x)}}\\
\langle R(y)\rangle &=& \frac{\int
d^{4}y\sqrt{\bar{g}(y)}R(y)}{\int d^{4}y\sqrt{\bar{g}(y)}}
\end{eqnarray}
and similarly for $\langle L(x)\rangle$ and $\langle L(y)\rangle$.

Thus there is a global interaction between the universes $X$ and
$Y$: The integral over the whole history of the $Y$-universe
changes the vacuum energy density of the $X$-universe. Assuming
then that at late times the fields settle near the absolute
minimum of their potential we have:
\begin{eqnarray}
R_{\mu\nu}(x) - \frac{1}{2}g_{\mu\nu}R(x) &=& -8\pi
Gg_{\mu\nu}\left[V(\bar{\phi_0}) - V(\phi_0)\right] -\frac{1}{2}
g_{\mu\nu}R(y)\\
R_{\alpha\beta}(y) - \frac{1}{2}\bar{g}_{\alpha\beta}R(y) &=&
-8\pi Gg_{\alpha\beta}\left[V(\phi_0) - V(\bar{\phi_0})\right] -
\frac{1}{2}g_{\alpha\beta}R(x).
\end{eqnarray}
Thus at late stages the effective cosmological constant vanishes:
\begin{equation}
R(x) = -R(y) = \frac{32}{3}\pi G\left[V(\phi_0) -
V(\bar{\phi_0})\right] = 0,
\end{equation}
since because of the antipodal symmetry $\phi_0 = \bar{\phi_0}$
and $V(\phi_0) = V(\bar{\phi_0})$.

This could also be seen as a back-reaction mechanism, from one
universe at the other.

\subsection{Duality Transformations}

\subsubsection{S-Duality}

A different proposal was considered in \cite{Ellwanger2004}, where
S-duality acting on the gravitational field is assumed to mix
gravitational and matter degrees of freedom. The purpose is to
show that whereas the original metric may be (A)dS, de dual will
be flat. It is assumed that:
\begin{equation}\label{asssd1}
R^{b}_{\; a}\equiv R^{ca}_{\;\; bc}=\Lambda\delta^{a}_{\; b},
\end{equation}
with $\Lambda$ the cosmological constant. The mixing between
gravitational and matter degrees of freedom is obtained through a
new definition of the gravitational dual of the Riemann tensor,
including the field strength $F_{abcd}$ of a 3-form field
$A_{abc}$. which equation of motion is simply
$F_{abcd}=\omega\epsilon_{abcd}$, with $\omega$ some constant, see
also section (\ref{Hawking}):
\begin{eqnarray}
\tilde{R}_{abcd}&=&\frac{1}{2}\epsilon_{abef}\left(R^{ef}_{\;\;
cd}
+ F^{ef}_{\;\; cd}\right) + \frac{1}{12}\epsilon_{abcd}, \nonumber\\
\tilde{F}_{abcd}&=& -\frac{1}{2}\epsilon_{abcd}R
\end{eqnarray}
such that:
\begin{eqnarray}
\tilde{\tilde{R}}_{abcd}&=& -R_{abcd}\nonumber\\
\tilde{\tilde{F}}_{abcd}&=& -F_{abcd}.
\end{eqnarray}
The equations of motion for the dual tensors become:
\begin{eqnarray}
\tilde{R}^{a}_{\; b}&=& 3\omega\delta^{a}_{\; b}\nonumber\\
\tilde{F}_{abcd}&=&
-\frac{1}{3}\Lambda\epsilon_{abcd}\equiv\tilde{\omega}\epsilon_{abcd}.
\end{eqnarray}
Therefore it seems that if the vev $\omega$ would vanish, the dual
Ricci tensor, in casu the dual cosmological constant would also
vanish. Hence the conclusion is that if we would `see' the dual
metric, determined by the dual Riemann tensor, we would observe a
flat spacetime.

However, with assumption (\ref{asssd1}), the trace of the
left-hand-side of Einstein's equation vanishes by definition.
Hence, also the trace of the energy-momentum tensor should vanish,
which in general is not the case. The field equations therefore
appear to be inconsistent with the above assumption, unless
$\omega =0$, which makes the addition of the field strength term
useless. This scenario, even aside from the other assumptions,
therefore cannot work.

Note that S-duality is an important concept in stringtheory. If
theories A and B are S-dual then $f_A(\alpha) = f_B(1/\alpha)$. It
relates type I stringtheory to the $SO(32)$ heterotic theory, and
type IIB theory to itself.

\subsubsection{Hodge Duality}

This duality between a $r$-form and a $(D-r)$-form in $D$
dimensions is studied \cite{NishinoRajpoot2004}, where the
cosmological constant is taken to be represented by a 0-form field
strength, which is just a constant. This is related to the
unimodular approach of section (\ref{unimodular}) in the sense
that they try to introduce the cosmological constant in a
different way in the Einstein-Hilbert action.

\subsection{Summary}

A symmetry principle as explanation for the smallness of the
cosmological constant in itself is very attractive. A viable
mechanism that sets the cosmological constant to zero would be
great progress, even if $\Lambda$ would turn out to be nonzero.
Since supersymmetry does not really seem to help, especially some
form of scale invariance stands out as a serious option. Needless
to say, it is hard to imagine how scale invariance could be used,
knowing that the world around us is not scale invariant. Particle
masses are small, but many orders of magnitude larger than the
observed cosmological constant.

Another option might be that a symmetry condition enforcing
$\rho_{vac}$ equal to zero, could be reflected in a certain choice
of boundary conditions. In such a scenario, the vacuum state would
satisfy different boundary conditions then excited states. The
$x\rightarrow ix$ transformation of section (\ref{imaginaryspace})
could be an example of this.

\section{Type II: Back-Reaction Mechanisms}

In this approach it is argued that any cosmological constant will
be automatically cancelled, or screened, to a very small value by
back-reaction effects on an expanding space. The effective
cosmological constant then is small, simply because the universe
is rather old. Often these effects are studied in an inflationary
background, where a cosmological constant is most dominant. The
physical idea of this mechanism can be understood in the context
of the energy-time uncertainty principle. For a particle with mass
$m$ and co-moving wavevector $\mathbf{k}$ in a spacetime with
scalefactor $a(t)$ we have:
\begin{equation}
E(\mathbf{k}, t)=\sqrt{m^2 + \|\mathbf{k}\|^2/a^2(t)}.
\end{equation}
Thus growth of $a(t)$ increases the time a virtual particle of
fixed $m$ and $\mathbf{k}$ can exist and, during inflation,
virtual particles with zero mass and long enough wavelength can
exist forever. The rate ($\Gamma$) at which they emerge from the
inflationary vacuum depends upon the type of particle. Most
massless particles are conformally invariant. In that case,
$\Gamma$ gives the number of particles emerging from the vacuum
per unit conformal time $\eta$, so the number per unit physical
time is:
\begin{equation}
\frac{dn}{dt} = \frac{d\eta}{dt}\frac{dn}{d\eta} =
\frac{\Gamma}{a}.
\end{equation}
Their emergence rate thus falls like $1/a(t)$. This means that
although those that are produced can exist forever, only very few
are created, and their total effect during inflation is
negligible, see e.g. \cite{prokowood2003}.

However, two familiar massless particles are not conformally
invariant, massless minimally coupled scalars and gravitons.
Therefore in these two sections we consider their effects in more
detail.

It should be noted that there exists a no-go theorem, derived by
Weinberg, see \cite{Weinbergreview} for details. The theorem
states that the vacuum energy density cannot be cancelled
dynamically, using a scalar field, without fine-tuning in any
effective four-dimensional theory with constant fields at late
times, that satisfies the following conditions:
\begin{enumerate}
\item General Covariance;
\item Conventional four-dimensional gravity is mediated by a
\textit{massless} graviton;
\item Theory contains a finite number of fields below the cutoff
scale;
\item Theory contains no negative norm states.
\end {enumerate}
Under these rather general assumptions the theorem states that the
potential for the compensator field, which should adjust the
vacuum energy to zero, has a runaway behavior. This means that
there is no stationary point for the potential of the scalar field
that should realize the adjustment, and thus the mechanism cannot
work.

\subsection{Scalar Field, Instabilities \textit{in} dS-Space}

The first attempts to dynamically cancel a 'bare' cosmological
constant were made by referring to instabilities in the case of a
scalar field in de Sitter space. A massless minimally coupled
scalar field $\phi$ has no de Sitter-invariant vacuum state and
the expectation value of $\phi^2$ is time-dependent. However, this
breaking of de Sitter invariance is not reflected by the
energy-momentum tensor, since $T_{\mu\nu}$ only contains
derivatives and hence is not sensitive to long-wavelength modes.
This changes if one includes interactions. Consider for example a
$\lambda\phi^4$. Then:
\begin{equation}
\langle
T_{\mu\nu}\rangle\sim\lambda\langle\phi^2\rangle^2g_{\mu\nu}\propto
t^2.
\end{equation}
So in this case it is possible for $\langle T_{\mu\nu}\rangle$ to
grow for some time, until higher order contributions become
important. The infrared divergence results in a mass for the field
which in turn stops the growth of $\langle T_{\mu\nu}\rangle$, see
for example \cite{Ford1985, Ford}.

Another illustrative, but unsuccessful attempt has been given by
Dolgov \cite{Dolgov82}. He used a rather simple classical model
for back-reaction:
\begin{equation}
\mathcal{L} =
\frac{1}{2}\left(\partial_{\alpha}\phi\partial^{\alpha}\phi - \xi
R\phi^2\right),
\end{equation}
where $R$ is the scalar curvature and $\xi$ a \textit{negative}
constant. The scalar field energy-momentum tensor at late times
approaches the form of a cosmological constant term:
\begin{equation}
8\pi G\langle T_{\mu\nu}\rangle\sim\Lambda_0g_{\mu\nu} +
\mathcal{O}(t^{-2}).
\end{equation}
$\Lambda_0=3H^2$ stands for the effective value of the
cosmological constant during a de Sitter phase so the leading
back-reaction term cancels this effect. The kinetic energy of the
growing $\phi$-field acts to cancel the cosmological constant. The
no-go theorem of the previous section is circumvented, since the
scalar field is not constant at late times.

Unfortunately, not only the cosmological constant term is driven
to zero, Newton's constant is also screened:
\begin{equation}
G_{eff} = \frac{G_0}{1+8\pi G|\xi|\phi^2}\sim\frac{1}{t^2},
\end{equation}
where $G_0$ is the ``bare" value of $G$ at times where $\phi=0$.
This is a fatal flaw of many of such approaches.

Other models of this kind were also studied by Dolgov, see
\cite{Dolgov2,Dolgov3,Dolgov2004} but these proved to be unstable,
leading quickly to a catastrophic cosmic singularity.

As we discussed, Weinberg's no-go theorem is widely applicable to
such screening mechanisms. However, it was noted in e.g.
\cite{Peccei:1987mm}, that conformal anomalies might provide a way
around this. The Lagrangian obtains an additional term
proportional to $\sqrt{g}\phi\Theta^{\mu}_{\mu}$, where
$\Theta^{\mu}_{\mu}$ is the effect of the conformal anomaly.

However, as already noted by Weinberg \cite{Weinbergreview}, this
does not provide a loophole to get around the no-go theorem. The
reason is that, although the field equation for $\phi$ now looks
like:
\begin{equation}
\frac{\partial\mathcal{L}}{\partial\phi}=\sqrt{-g}\left(T^{\mu}_{\mu}
+ \Theta^{\mu}_{\mu}\right),
\end{equation}
which may suggest an equilibrium value for $\phi$ with zero trace,
this is not sufficient for a flat space solution. The Einstein
equation for a constant metric now becomes:
\begin{equation}
0=\frac{\partial\mathcal{L}_{eff}}{\partial g_{\mu\nu}}\propto
e^{2\phi}\mathcal{L}_0 + \phi\Theta^{\mu}_{\mu},
\end{equation}
and the extra factor of $\phi$ shows that these two conditions are
not the same. The reason is that the term $\Theta^{\mu}_{\mu}$
does not simply end up in $T^{\mu}_{\mu}$.

\subsubsection{Radiative Stability in Scalar Field Feedback
Mechanism}

Another approach deserves to be mentioned here. This concerns a
model that does not solve the cosmological constant problem, but
does seem to provide a way to protect a zero or small cosmological
constant against radiative corrections, without using a symmetry,
\cite{MukohyamaRandall2003,Mukohyama2003}. This is achieved using
a scalar field with a non-standard, curvature dependent kinetic
term, such that in the limit where the scalar curvature goes to
zero, the kinetic term vanishes.
\begin{eqnarray}
S &=& \int d^4x\sqrt{-g}\left(\frac{R}{2\kappa^2} + \alpha R^2 +
L_{kin} - V(\phi)\right)\nonumber\\
L_{kin} &=& \frac{\kappa^{-4}K^q}{2qf^{2q-1}},
\end{eqnarray}
where $q$ is a constant that has to be $q>1/2$ for stability
reasons, and $f$ is a function of the scalar curvature $R$,
postulated to vanish at $R=0$ and that behaves near $R=0$ as:
\begin{equation}
f(R)\sim\left(\kappa^4R^2\right)^m,
\end{equation}
with $\kappa$ the Planck length. The parameter $\alpha$ is assumed
to be $\alpha >0$ to stabilize gravity at low energies, $m$ is an
integer that satisfies $2(m-1)>q(2q-1)$ and $K\equiv
-\kappa^4\partial^{\mu}\phi\partial_{\mu}\phi$.

The true value of the vacuum energy in this approach is not zero,
but the peculiar dynamics makes the universe settle down to a near
zero energy state. The scalar field stops rolling and its kinetic
terms diverges.

The two main problems with this scenario are: 1) This specific
kinetic term is chosen by hand, not motivated by a more
fundamental theory, 2) all other fields settle to their ground
state faster than the vacuum energy, making the universe empty,
and reheating necessary, to thermally populate the universe again.

Other models where some dynamical feedback mechanism is proposed
based on a non-standard kinetic term can be found in
\cite{Rubakov1999,HebeckerWetterich2000,Hebecker2001,Wetterich2002}.
An interesting conjecture is made on the existence of a conformal
fixed point, possibly related to dilatation symmetry
\cite{Wetterich1987}. However, these models still need
fine-tuning, and it is unclear whether they are experimentally
viable, see \cite{Mukohyama2003}.

\subsection{Dilaton}

A natural scalar field candidate to screen the cosmological
constant could be the dilaton, which appears in string theory an
compactified supergravity theories. In the presence of a dilaton,
all mass scales arise multiplied with an exponential:
\begin{equation}
V_0(\phi)\sim M^4e^{4\lambda\phi},
\end{equation}
with $\phi$ the dilaton, and $\lambda$ a coupling constant. The
minimum of this obtained for the value $\phi_0 = -\infty$, which s
known as the `dilaton runaway problem': couplings depend typically
on $\phi$, and these tend to go to zero, or infinity sometimes, in
this limit. Moreover, all mass scales have this similar scaling
behavior, so particle masses also vanish. Besides, the dilaton
itself is nearly massless when it reaches the minimum of its
potential, leading to long-range interactions that are severely
constrained. Note that quintessence ideas can only be maintained
as long as the new hypothetical scalar particle does not couple to
the standard model fields, contrary to the dilaton.

In summary, the dynamical cancellation of a cosmological constant
term by back-reaction effects of scalar fields is hard to realize.
Let's focus therefore on a purely gravitational back-reaction
mechanism.

\subsection{Gravitons, Instabilities \textit{of} dS-Space}

Gravitational waves propagating in some background spacetime
affect the dynamics of this background. This back-reaction can be
described by an effective energy-momentum tensor $\tau_{\mu\nu}$.

\subsubsection{Scalar-type Perturbations}

In \cite{Brandenberger, Brandenberger1} the back-reaction for
scalar gravitational perturbations is studied. It is argued this
might give a solution to the CC problem.

At linear order, all Fourier modes of the fluctuations evolve
independently. However, since the Einstein equations are
non-linear, retaining higher order terms in the perturbation
amplitude leads to interactions between the different perturbation
modes: they define a gravitational back-reaction.

The idea is first to expand Einstein equations to second order in
the perturbations, then to assume that linear terms satisfy
equations of motion (and hence cancel). Next the spatial average
is taken of the remaining terms and the resulting equations are
regarded as equations for a new homogeneous metric
$g_{\mu\nu}^{(0,br)}$, where the superscript $(0,br)$ denotes
first the order in perturbation theory and the fact that
back-reaction is taken into account:
\begin{equation}
G_{\mu\nu}\left(g_{\alpha\beta}^{(0,br)}\right) = -8\pi
G\left[T_{\mu\nu}^{(0)} + \tau_{\mu\nu}\right]
\end{equation}
and $\tau_{\mu\nu}$ contains terms resulting from averaging of the
second order metric and matter perturbations:
\begin{equation}
\tau_{\mu\nu} = \langle T_{\mu\nu}^{(2)} - \frac{1}{8\pi
G}G_{\mu\nu}^{(2)}\rangle.
\end{equation}
In other words, the first-order perturbations are regarded as
contributing an extra energy-momentum tensor to the zeroth-order
equations of motion; the effective energy-momentum tensor of the
first-order equations renormalizes the zeroth-order
energy-momentum tensor. This is a somewhat tricky approach and it
is not clear whether one can consistently derive the equations of
motion in this way, see for example \cite{Grishchuk1994,
MartinSchwarz1997,Mukhanovetal1996,Abramoetal1997,Grishchuk1998,MartinSchwarz1998,
Unruh1998}.

Now work in longitudinal gauge and take the matter to be described
by a single scalar field for simplicity. Then there is only one
independent metric perturbation variable denoted $\phi(x,t)$. The
perturbed metric is:
\begin{equation}
ds^{2} = (1 + 2\phi)dt^{2} - a(t)^{2}(1 -
2\phi)\delta_{ij}dx^{i}dx^{j}.
\end{equation}
Calculating the $\tau_{00}$ and $\tau_{ij}$ elements and using
relations valid for the period of inflation, Brandenberger's main
result is that the equation of state of the dominant infrared
contribution to the energy-momentum tensor $\tau_{\mu\nu}$ which
describes back-reaction, takes the form of a negative CC:
\begin{equation}
p_{br} = -\rho_{br}, \quad\quad\quad \rho_{br} < 0.
\end{equation}
This leads to the speculation that gravitational back-reaction may
lead to a dynamical cancellation mechanism for a bare CC since
$\tau_{0}^0 \propto\langle\phi^2\rangle$, which is proportional to
IR phase space and this diverges in a De Sitter universe. Long
wavelength modes are those with wavelength longer than $H$, and as
more and more modes cross the horizon, $\langle\phi^2\rangle$
grows. To end inflation this way, however, takes an enormous
number of e-folds, see \cite{Martineau:2005zu} for a recent
discussion.

However, as pointed out in \cite{Unruh1998}, the spatially
averaged metric is not a local physical observable: averaging over
a fixed time slice, the averaged value of the expansion will not
be the same as the expansion rate at the averaged value of time,
because of the non-linear nature of the expansion with time. In
other words, locally this `achieved renormalization', i.e. the
effect of the perturbations, is identical to a coordinate
transformation of the background equations and not a physical
effect. A similar conclusion was obtained in
\cite{KodamaHamazaki1997,AbramoWoodard2001}.

Brandenberger and co-workers have subsequently tried to improve
their analysis by identifying a local physical variable which
describes the expansion rate
\cite{BrandenbergerGeshnizjani2003,Brandenberger2}. This amounts
to adding another scalar field that acts as an independent
physical clock. Within this procedure they argue that
back-reaction effects are still significant in renormalizing the
cosmological constant.

It is however far from clear whether this scenario is consistent
and whether the effects indeed are physical effects. One of the
main points is that by performing a coordinate transformation, one
can locally always find coordinates such that at a given point
$P$, $g'_{\mu\nu}(x'_P)=\eta_{\mu\nu}$ and $\partial
g'_{\mu\nu}/\partial x'_{\alpha}=0$ evaluated at $x=x_P$, simply
constructing a local inertial frame at the point $P$. The second
and higher order derivatives of the metric can of course not be
made to vanish and measure the curvature. The perturbations are
small enough that we do not notice any deviation from homogeneity
and isotropy, but are argued to be large enough to alter the
dynamics of our universe, which sounds contradictory. In
\cite{Ishibashi:2005sj} especially, on general grounds these
effects are argued to be unphysical and therefore cannot provide a
solution to the cosmological constant problem.

Besides, this build-up of infrared scalar metric perturbations
(vacuum fluctuations, stretched beyond the Hubble-radius) is set
in an inflationary background and since the individual effects are
extremely weak a large phase-space of IR-modes, i.e. a long period
of inflation, is needed. The influence on today's cosmological
constant is unclear.

\subsubsection{Long-Wavelength Back-Reaction in Pure
Gravity}\label{TsamisWoodard}

Closely related are studies by Tsamis and Woodard, see
\cite{TsamisWoodard0, TsamisWoodard1, TsamisWoodard2,
TsamisWoodard3, TsamisWoodard4, TsamisWoodard5, TsamisWoodard6}
concerning the back-reaction of long-wavelength gravitational
waves in pure gravity with a bare cosmological constant. Leading
infrared effects in quantum gravity are, contrary to what is often
assumed, similar to those of QED, see \cite{Weinberg1965}.

When $\Lambda\neq 0$, the lowest dimensional self-interaction term
is of dimension three, a three-point vertex with no derivatives
(corresponding to the $\Lambda\sqrt{-g}$-term). The IR behavior of
the theory with cosmological constant is therefore very different
from that without. Tsamis and Woodard christen it Quantum
Cosmological Gravity, or QCG for short, and study it on an
inflationary background. Here the infrared divergences are
enhanced: since the spatial coordinates are exponentially expanded
with increasing time, their Fourier conjugates, the spatial
momenta, are redshifted to zero. The IR effects originate from the
low end of the momentum spectrum, so they are strengthened when
this sector is more densely populated.

Since other particles are either massive, in which case they
decouple from the infrared, or conformally invariant, and
therefore do not feel the de Sitter redshift, gravitons must
completely dominate the far IR. The typical strength of quantum
gravitational effects during inflation at scale $M$ is:
\begin{equation}
G\Lambda=8\pi\left(\frac{M}{M_{P}}\right)^4,
\end{equation}
which for GUT-scale inflation becomes $G\Lambda=10^{-11}$ and for
electroweak-scale inflation $G\Lambda=10^{-67}$.

The classical background in conformal coordinates is:
\begin{eqnarray}
-dt^2+e^{2Ht}d\mathbf{x}\cdot d\mathbf{x}&=&\Omega^2\left(-du^2+
d\mathbf{x}\cdot d\mathbf{x}\right)\\
\Omega\equiv\frac{1}{Hu}=\exp(Ht)
\end{eqnarray}
and $H^2\equiv\frac{1}{3}\Lambda$. For convenience, perturbation
theory is formulated in terms of a pseudo-graviton field
$\psi_{\mu\nu}$:
\begin{equation}
g_{\mu\nu}\equiv\Omega^2\tilde{g}_{\mu\nu}\equiv\Omega^2(\eta_{\mu\nu}
+ \kappa\psi_{\mu\nu})
\end{equation}
where $\kappa^2\equiv 16\pi G$.

Because of homogeneity and isotropy of the dynamics and the
initial state, the amputated 1-point function, can be written in
terms of two functions of conformal time $u$:
\begin{equation}
D_{\mu\nu}^{\rho\sigma}\langle0|\kappa\psi_{\rho\sigma}(x)|0\rangle
= a(u)\bar{\eta}_{\mu\nu} + c(u)\delta^{0}_{\mu}\delta^{0}_{\nu},
\end{equation}
where $D_{\mu\nu}^{\rho\sigma}$ is the gauge fixed kinetic
operator, and a bar on $\eta_{\mu\nu}$ indicates that temporal
components of this tensor are deleted:
\begin{equation}
\eta_{\mu\nu} =
\bar{\eta}_{\mu\nu}\delta_{\mu}^{(0)}\delta_{\nu}^{(0)}.
\end{equation}
The pseudo-graviton kinetic operator $D_{\mu\nu}^{\rho\sigma}$
splits in two parts, a term proportional to
$D_A\equiv\Omega(\partial^2+\frac{2}{u^2})\Omega$, which is the
kinetic operator for a massless minimally coupled scalar, and a
part proportional to $D_C\equiv\Omega\partial^2\Omega$, the
kinetic operator for a conformally coupled scalar.

After attaching the external legs one obtains the full 1-point
function, which has the same form, but with different components:
\begin{equation}
\langle0|\kappa\psi_{\rho\sigma}(x)|0\rangle =
A(u)\bar{\eta}_{\mu\nu} + C(u)\delta_{\mu}^{0}\delta_{\nu}^{0}.
\end{equation}
The functions $A(u)$ and $C(u)$ obey the following differential
equations:
\begin{eqnarray}
-\frac{1}{4}D_A\left[A(u)-C(u)\right]&=& a(u)\nonumber\\
D_CC(u)&=& 3a(u)+c(u)
\end{eqnarray}
The functions $a(u)$ and $A(u)$ on the one hand, and $c(u)$ and
$C(u)$ on the other, are therefore related by retarded Green's
functions $G_{A,C}^{ret}$ for the massless minimally coupled and
conformally coupled scalars:
\begin{eqnarray}
A(u) &=& -4G_{A}^{ret}[a](u) + G_{C}^{ret}[3a+c](u),\nonumber\\
C(u)&=&G_{C}^{ret}[3a+c](u)
\end{eqnarray}
In terms of these functions $A(u)$ and $C(u)$ the invariant
element in comoving coordinates reads:
\begin{equation}
\hat{g}_{\mu\nu}(t,\mathbf{x})dx^{\mu}dx^{\nu}=-\Omega^2\left[1-C(u)\right]du^2+\Omega^2\left[1+A(u)\right]d\mathbf{x}\cdot
d\mathbf{x}.
\end{equation}
This gives the following identification:
\begin{eqnarray}\label{idenTW}
R(t)&=&\Omega\sqrt{1+A(u)},\nonumber\\
d(t)&=&-\Omega\sqrt{1-C(u)}du,\quad\quad\mbox{and}\quad\quad
d(Ht)=-\sqrt{1-C(u)}d[\ln(Hu)]
\end{eqnarray}
Using this we can find the time dependence of the effective Hubble
parameter:
\begin{equation}
H_{eff}(t)=\frac{d}{dt}\ln\left(R(t)\right)=
\frac{H}{\sqrt{1-C(u)}}\left(1-\frac{\frac{1}{2}u\frac{d}{du}A(u)}{1+A(u)}\right).
\end{equation}

The backreaction of the IR gravitons therefore acts to screen the
bare cosmological constant, originally present. The improved
results\footnote{Papers before 1997 yield different results.} in
terms of:
\begin{equation}
\epsilon\equiv\left(\frac{\kappa
H}{4\pi}\right)^2=\frac{G\Lambda}{3\pi}=\frac{8}{3}\left(\frac{M}{M_P}\right)^4
\end{equation}
turn out to be:
\begin{eqnarray}
A(u)&=&\epsilon^2\left\{\frac{172}{9}\ln^3(Hu) +
(\mbox{subleading})\right\} + \mathcal{O}(\epsilon^3),\\
C(u)&=&\epsilon^2\left\{57\ln^2(Hu) + (\mbox{subleading})\right\}
+ \mathcal{O}(\epsilon^3)
\end{eqnarray}
Using (\ref{idenTW}) we find:
\begin{equation}
Ht = -\left\{1-\frac{19}{2}\epsilon^2\ln^2(Hu) +
\dots\right\}\ln(Hu)
\end{equation}
This implies that $\ln(Hu)\approx -Ht$ to very good approximation,
therefore $A(u)$ can be written:
\begin{equation}
A(u)=-\frac{172}{9}\epsilon^2(Ht)^3+\ldots
\end{equation}
and we arrive at:
\begin{eqnarray}
H_{eff}(t)&\approx& H + \frac{1}{2}\frac{d}{dt}\ln(1+A),\nonumber\\
&\approx&
H\left\{1-\frac{\frac{86}{3}\epsilon^2\left(Ht\right)^2}{1-\frac{172}{9}\epsilon^2\left(Ht\right)^3}\right\}
\end{eqnarray}
The induced energy density, which acts to screen the original
cosmological constant present gives:
\begin{eqnarray}
\rho(t)&\approx&\frac{\Lambda}{8\pi
G}\left\{-\frac{1}{H}\frac{\dot{A}}{1+A} +
\frac{1}{4H^2}\left(\frac{\dot{A}}{1+A}\right)^2\right\}\nonumber\\
&\approx&\frac{\Lambda}{8\pi
G}\left\{-\frac{\frac{172}{3}\epsilon^2\left(Ht\right)^2}{1-\frac{172}{9}\epsilon^2\left(Ht\right)^3}
+
\left(\frac{\frac{86}{3}\epsilon^2\left(Ht\right)^2}{1-\frac{172}{9}\epsilon^2\left(Ht\right)^3}\right)^2\right\}
\end{eqnarray}
This can be written more intuitively, to better see the magnitude
of the effect as follows:
\begin{equation}
H_{eff}(t)=H\left\{1-\epsilon^2\left[\frac{1}{6}(Ht)^2 +
(\mbox{subleading})\right] +\mathcal{O}(\kappa^6)\right\}
\end{equation}
and the induced energy density and pressure, in powers of $H$:
\begin{eqnarray}
\rho(t)&=&\frac{\Lambda}{8\pi G} + \frac{(\kappa
H)H^4}{2^6\pi^4}\left\{-\frac{1}{2}\ln^2a + \mathcal{O}(\ln
a)\right\} + \mathcal{O}(\kappa^4)\nonumber\\
p(t)&=&-\frac{\Lambda}{8\pi G} + \frac{(\kappa
H)H^4}{2^6\pi^4}\left\{\frac{1}{2}\ln^2a + \mathcal{O}(\ln
a)\right\} + \mathcal{O}(\kappa^4).
\end{eqnarray}

The number of e-foldings needed to make the backreaction effect
large enough to even end inflation is:
\begin{equation}
N\sim\left(\frac{9}{172}\right)^{\frac{1}{3}}\left(\frac{3\pi}{G\Lambda}\right)^{\frac{2}{3}}
=
\left(\frac{81}{11008}\right)^\frac{1}{3}\left(\frac{M_P}{M}\right)^\frac{8}{3}
\end{equation}
where $M$ is the mass scale at inflation and $M_P$ the Planck
mass. For inflation at the GUT scale this gives $N\sim 10^7$
e-foldings. This enormously long period of inflation, much longer
than in typical inflation models, is a direct consequence of the
fact that gravity is such a weak interaction.

In other words, the effect might be strong enough to effectively
kill any cosmological constant present, as long as such a long
period of inflation is acceptable. There do exist arguments that
the number of e-folds is limited to some~85, see
\cite{BanksFischler2003} for details, but these are far from
established. Another issue is that these results have been
obtained for a very large cosmological constant during inflation.
It is unclear what this means for the present day vacuum energy of
the universe. Perturbative techniques break down when the effect
becomes too strong, making this difficult to answer.

This breaking however is rather soft, since each elementary
interaction remains weak. Furthermore, a technique following
Starobinski \cite{StarobinskyYokoyama1994} is used in which
non-perturbative aspects are absorbed in a stochastic background
that obeys the classical field equations \cite{TsamisWoodard5}.

It is then argued \cite{TsamisWoodard5} that eventually the
screening must overcompensate the original bare cosmological
constant, leading to a period of deflation. This happens because
the screening at any point derives from a coherent superposition
of interactions from within the past lightcone and the invariant
volume of the past lightcone grows faster as the expansion slows
down. Now thermal gravitons are produced that act as a thermal
barrier, that grows hotter and denser as deflation proceeds.
Incoming virtual IR modes scatter off this barrier putting a halt
to the screening process. The barrier dilutes and the expansion
takes over again.

However, discussions are still going on, debating whether these
screening effects are real physical effects, or gauge artifacts,
see \cite{Abramo:2001dd,Woodard2004b,Ishibashi:2005sj}.
Especially, the argued cumulative nature of the effect, makes it
hard to understand how local physics is affected.

Another objection may be raised in that throughout the above
calculation, the `primordial' cosmological constant $\Lambda$ was
used. The mechanism, however, screens the cosmological constant,
which implies that the effective cosmological constant should be
used instead. The strength of the effect would then be even
weaker, since this is controlled by $G\Lambda$. A much larger
number of e-folds would then be necessary to stop inflation.

\subsection{Screening as a Consequence of the Trace Anomaly}

In
\cite{Tomboulis1988,AntoniadisMottola1991,AntoniadisMottolaMazur98}
it is argued that the quantum effects of the trace anomaly of
massless conformal fields in 4 dimensions leads to a screening of
the cosmological constant. The effective action of 4D gravity
yields an extra new spin-0 degree of freedom in the conformal
sector, or trace of the metric. At very large distance scales this
trace anomaly induced action dominates the standard Einstein
action and gives an IR fixed point where scale invariance is
restored.

The idea is similar to that in the previous section,
(\ref{TsamisWoodard}). One tries to find a renormalization group
screening of the cosmological constant in the IR, but instead of
taking full quantum gravity effects, only quantum effects of the
conformal factor are considered. See also \cite{Odintsov:1990rs}
for a related earlier study.

The authors conclude that the effective cosmological constant and
inverse Newton's constant in units of Planck mass decreases at
large distances and that $G_N\Lambda\rightarrow 0$ at the IR fixed
point in the infinite volume limit.

However, the cosmological constant problem manifests itself
already at much smaller distances and moreover, it is unclear
whether this scenario is compatible with standard cosmological
observations. Moreover, like the other approaches in this chapter,
it relies heavily on quantum effects having a large impact at
enormous distance scales. As argued in the previous section, it is
debatable whether these effects can be sufficiently significant.

\subsection{Running $\Lambda$ from Renormalization Group}\label{RenormalizationGroup}

In
\cite{Shapiro1994,Elizalde1994,ShapiroSola2000,ShapiroSola2004,Babic2004,ReuterWeyer2004}
a related screening of the cosmological constant is studied,
viewing $\Lambda$ as a parameter subject to renormalization group
running. The cosmological constant than becomes a scaling
parameter $\Lambda(\mu)$, where $\mu$ is often identified with the
Hubble parameter at the corresponding epoch, in order to make the
running of $\Lambda$ smooth enough to agree with all existing
data, \cite{Espana-BonetLapuenteShapiroSola}.

However, renormalization group equations generally give
logarithmic corrections:
\begin{equation}
\mu\frac{d\lambda}{d\mu}=A_0,\quad\quad\rightarrow\quad\quad
\frac{\lambda(\mu)}{\lambda_0}= 1-q_1\ln\frac{\mu}{\mu_0},
\end{equation}
where $q_1\sim m^4/\lambda_0$, which makes it hard to see how this
can ever account for the suppression of a factor of $10^{120}$
needed for the cosmological constant. In the above refs., a
different running is considered:
\begin{equation}
\mu\frac{d\lambda}{d\mu}=A_1\mu^2,\quad\quad\rightarrow\quad\quad
\frac{\lambda(\mu)}{\lambda_0}= L_0 + L_1\frac{\mu^2}{\mu_{0}^2},
\end{equation}
where the running is still very small, since $L_1\sim
m^2/M_{P}^2$.

Although this running is very slow, it could possibly be measured
as a quintessence of phantom dark energy and be consistent with
all data, as long as $0\leq |\nu| \ll 1$ \cite{Sola:2005et}. As a
solution to the cosmological constant problem, it obviously cannot
help.

In refs. \cite{ReuterWeyer2003,ReuterWeyer2004}, it is argued that
there may be a UV fixed point at which gravity becomes
asymptotically free. If there would be an IR fixed point at which
$\Lambda_{eff}=0$ this could shed some new light on the
cosmological constant problem. This scaling also effects $G$,
making it larger at larger distances.

\subsubsection{Triviality as in $\lambda\phi^4$ Theory}

The Einstein Hilbert action with a cosmological constant can be
rewritten as \cite{JackiwNunezPi}:
\begin{equation}
S=-\frac{3}{4\pi}\int
d^4x\sqrt{-\hat{g}}\left(\frac{1}{12}R(\hat{g})\phi^2 +
\frac{1}{2}\hat{g}^{\mu\nu}\partial_{\mu}\phi\partial_{\nu}\phi
-\frac{\lambda}{4!}\phi^4\right)
\end{equation}
after rescaling the metric tensor as:
\begin{equation}
g_{\mu\nu}=\varphi^2\hat{g}_{\mu\nu}, \quad\quad
ds^2=\varphi^2\hat{d}s^2
\end{equation}
and defining:
\begin{equation}
\phi = \frac{\varphi}{\sqrt{G}}, \quad\quad
\Lambda=\frac{\lambda}{4G}.
\end{equation}
Now it is suggested that the same arguments first given by Wilson
\cite{Wilson1971}, that are valid in ordinary
$\lambda\phi^4$-theory, might also hold here and that this term is
suppressed quantum mechanically.

It is noted that perturbative running as in normal
$\lambda\phi^4$-theory is by far not sufficient, but the idea is
that perhaps there might be some non-perturbative suppression.
Similar ideas have been contemplated by Polyakov,
\cite{Polyakov2000}.

\subsection{Summary}

Finding a viable mechanism that screens the original possibly
large cosmological constant to its small value today, is a very
difficult task. Weinberg's no-go theorem puts severe limits on
this approach. Back-reaction effects, moreover, are generally
either very weak, or lead to other troublesome features like a
screened Newton's constant.

The underlying idea however that the effective cosmological
constant is small simply because the universe is old, is
attractive and deserves full attention.

\section{Type III: Violating the Equivalence Principle}

An intriguing way to try to shed light on the cosmological
constant problem is to look for violations of the equivalence
principle of general relativity. The near zero cosmological
constant could be an indication that vacuum energy contrary to
ordinary matter-energy sources does not gravitate.

The approach is based not on trying to eliminate any vacuum
energy, but to make gravity numb for it. This requires a
modification of some of the building blocks of general relativity.
General covariance (and the absence of ghosts and tachyons)
requires that gravitons couple universally to all kinds of energy.
Moreover, this also fixes uniquely the low energy effective action
to be the Einstein-Hilbert action. If gravity were not mediated by
an exactly massless state, this universality would be avoided. One
might hope that vacuum energy would then decouple from gravity,
thereby eliminating the gravitational relevance of it and thus
eliminate the cosmological constant problem.

\subsection{Extra Dimensions, Braneworld Models}

Since the Casimir effect troubles our notion of a vacuum state,
the cosmological constant problem starts to appear when
considering distances smaller than a millimeter or so. Therefore,
extra dimensions with millimeter sizes might provide a mechanism
to understand almost zero 4D vacuum energy, since in these
scenarios gravity is changed at distances smaller than a
millimeter. This size really is a sort of turn-over scale. Somehow
all fluctuations with sizes between a Planck length and a
millimeter are cancelled or sum up to zero.

Besides, it is conceivable that the need to introduce a very small
cosmological constant or some other form of dark energy to explain
an accelerating universe nowadays, is a signal that general
relativity breaks down at very large distance scales. General
relativity however, works very well on scales from $10^{-1}$~mm to
at least $10^{14}$~cm, the size of the solar system. This puts
severe constraints on alternative theories. Extra-dimensional
models, like the early Kaluza-Klein scenarios, generically have
additional degrees of freedom, often scalar fields, that couple to
the four dimensional energy-momentum tensor and modify
four-dimensional gravity. A four dimensional massless graviton has
two physical degrees of freedom, a five dimensional one five, just
like a massive 4-dimensional graviton\footnote{In general, the
total number of independent components of a rank 2 symmetric
tensor in $D$ dimensions is $D(D+1)/2$, however, only $D(D-3)/2$
of those correspond to physical degrees of freedom of a
$D$-dimensional massless graviton; the remaining extra components
are the redundancy of manifestly gauge and Lorentz invariant
description of the theory.}. There are however, strong
experimental constraints on such scalar-tensor theories of
gravity, see for example \cite{Will:2001mx,Durrer:2003rg}.

A lot of research in this direction in recent years has been
devoted to braneworld models in $D=4+N$ dimensions, with $N$ extra
spatial dimensions. In this setting the cosmological constant
problem is at least as severe as in any other, but new mechanisms
of cancelling a vacuum energy can be thought of. The general idea
is that our world is confined on a hypersurface, a brane, embedded
in a higher dimensional spacetime. The standard model fields are
restricted to live on a 3-brane, while only gravitons can
propagate in the full higher dimensional space. To reproduce the
correct 4-dimensional gravity at large distances three approaches
are known. Usually one takes the extra dimensions to cover a
finite volume and compactifies the unseen dimensions. One of the
earliest approaches was by Rubakov and Shaposhnikov
\cite{RubakovShaposhnikov} who unsuccessfully tried to argue that
the 4D cosmological constant is zero, since 4D vacuum energy only
curves the extra dimensions.

In this chapter we will first briefly review the Randall-Sundrum
models and show why they cannot solve the cosmological constant
problem. Next we focus on the DGP-model with infinite volume extra
dimensions. This is a very interesting setup, but also a good
example of the difficulties one faces in deconstructing a higher
dimensional model to a viable 4D world meeting all the GR
constraints. A rather more speculative but perhaps also more
promising approach is subsequently discussed, in which Lorentz
invariance is spontaneously broken to yield a Higgs mechanism
analog for gravity. Before concluding with a summary, we discuss
yet another option, where one considers the graviton to be a
composite particle.

\subsubsection{Randall-Sundrum Models, Warped Extra
Dimensions}\label{section51}

There are in fact two different models known as Randall-Sundrum
models, dubbed RS-I and RS-II. We begin with RS-I.

This model consists of two 3-branes at some distance from each
other in the extra dimension. One brane, called the ``hidden
brane'' has positive tension, while the other one, the ``visible
brane'', on which we are supposed to live, has negative tension.
Both branes could have gauge theories living on them. All of the
Standard Model fields are localized on the brane, and only gravity
can propagate through the entire higher dimensional space.

The equation of motion looks as follows:
\begin{eqnarray}
&& M_{\ast}\sqrt{G}\left(R_{AB}-\frac{1}{2}G_{AB}R\right)-M_{\ast}\Lambda\sqrt{G}G_{AB}\nonumber\\
&=&
T_{hid}\sqrt{g_{hid}}g_{\mu\nu}^{hid}\delta_{A}^{\mu}\delta_{B}^{\nu}\delta(y)
+
T_{vis}\sqrt{g_{vis}}g_{\mu\nu}^{vis}\delta_{A}^{\mu}\delta_{B}^{\nu}\delta(y-y_0),
\end{eqnarray}
with notations:
\begin{equation}
g_{\mu\nu}^{hid}(x)=G_{\mu\nu}(x,y=0),\quad\
g_{\mu\nu}^{vis}(x)=G_{\mu\nu}(x,y=y_0).
\end{equation}
Furthermore, $M_{\ast}$ is the 5-dimensional Planck mass, which
has to satisfy $M_{\ast}\gtrsim 10^8$~GeV, in order not to spoil
Newtonian gravity at distances $l\lesssim 0.1$~mm.

The $y$-direction is compactified on an orbifold
$S_1/\mathbf{Z}_2$. With the above assumptions for the
brane-tensions and bulk CC, it can be shown that there exists the
following static solution, with a flat 4D-metric:
\begin{equation}
ds^2=e^{-|y|/L}\eta_{\mu\nu}dx^{\mu}dx^{\nu} + dy^2
\end{equation}
The minus sign in the exponential factor occurs because of the
assumption that our visible brane has a negative tension. As a
result of this `warp-factor', all masses on the visible brane are
suppressed, compared to their natural value. For the Higgs mass
for example, one obtains:
\begin{equation}
m^2=e^{-y_0/L}m_{0}^2
\end{equation}
a small hierarchy in $y_0/L$ therefore leads to a large hierarchy
between $m$ and $m_0$, which would solve the `ordinary' hierarchy
problem.

Moreover, despite the fact that the brane tension on the visible
brane is negative, it is possible that it still has a flat space
solution. Fine-tuning is necessary to obtain this result, and
besides, this solution is not unique. Other, non-flat space
solutions also exist. Therefore, this cannot help in solving the
cosmological constant problem, but it is interesting to see that a
4D cosmological constant can be made to curve only extra
dimensions.

Alternatively, the extra dimensions can be kept large,
uncompactified, but warped, as in the Randall-Sundrum type-II
models, in which there is only one brane. In this case the size of
the extra dimensions can be infinite, but their volume $\int
dy\sqrt{G}$, is still finite. The warp-factor causes the graviton
wavefunction to be peaked near the brane, or, in other words,
gravity is localized, such that at large 4D-distances ordinary
general relativity is recovered. The same bound as in RS-I applies
to the 5D Planck mass.

The action now reads:
\begin{equation}
S = \frac{1}{2}M_{\ast}^3\int
d^4x\int_{-\infty}^{+\infty}dy\sqrt{G}(R_5 - 2\Lambda_5)+\int
d^4x\sqrt{g}(\Lambda_4 + \mathcal{L}_{SM}),
\end{equation}
where $\Lambda_4$ denotes the 4D brane tension and $\Lambda_5$ the
bulk cosmological constant, which is assumed to be negative. The
equation of motion derived from this action, ignoring now
$\mathcal{L}$ is:
\begin{equation}
M_{\ast}\sqrt{G}\left(R_{AB}-\frac{1}{2}G_{AB}R\right) =
-M_{\ast}^3\Lambda_5\sqrt{G}G_{AB} +
\Lambda_4\sqrt{g}g_{\mu\nu}\delta_{A}^{\mu}\delta_{B}^{\nu}\delta(y),
\end{equation}
indicating that the brane is located at $y=0$. This equation has
the same flat space solution as above, but, again, at the expense
of fine-tuning $\Lambda_5$ and $\Lambda_4$.

Gravity in the $4D$ subspace reduces to GR up to some very small
Yukawa-type corrections. Unfortunately however, with regard to the
cosmological constant problem, the model suffers from the same
drawbacks as RS-I. All fundamental energy scales are at the TeV
level, but the vacuum energy density in our 4D-world is much
smaller.

\subsubsection{Self-Tuning Solutions}

Transmitting any contribution to the CC to the bulk parameters, in
such a way that a 4D-observer does not realize any change in the
4D geometry seems quite spurious. It would become more interesting
if this transmission would occur automatically, without the
necessity of re-tuning the bulk quantities by hand every time the
4D vacuum energy changes. Models that realize this are called
\textit{self-tuning} models (see for example \cite{Nilles}). A
severe drawback that all these models face is that this scenario
does not exclude `\textit{nearby curved solutions}'. This means
that in principle there could exist solutions for neighboring
values of some bulk parameters, which result in a curved 4D space.
Besides, there are additional problems such as a varying effective
Planck mass, or varying masses for fields on the brane. So far no
mechanism without these drawbacks has been found. See
\cite{Burgess1,Burgess2} for recent studies in favor of this
approach.

Another serious problem is that in many proposals, the 4D brane
tension creates a deficit angle in the bulk, which easily becomes
larger than $2\pi$. The cosmological constant problem rises again
in a different fine-tuning problem. For a recent review of this
approach and many references, see \cite{Burgess:2005wu}.

A related approach, considering a warped higher dimensional
geometry, is studied in refs.
\cite{Verlinde1999,Verlinde:1999mg,Schmidhuber1999,Schmidhuber2000}.
It is argued that once a cosmological constant vanishes in the UV,
there exist solutions such that it will not be regenerated along
the renormalization group flow. Any vacuum energy is cancelled by
a decreasing warp factor, ensuring a flat space solution on the
brane. However, these are not the solutions and there exists no
argument why they should be preferred. Note however, that this is
quite contrary to ordinary renormalization group behavior, as
studied in section \ref{RenormalizationGroup}.

\subsubsection{Infinite Volume Extra
Dimensions}\label{infiniteed}

In
\cite{Arkani-HamedDimopoulosDvaliGabadadze,DvaliGabadadzePorrati2000,DvaliGabadadze2000,DvaliGabadadzeShifman20021,DvaliGabadadzeShifman20022,Gabadadzereview},
a model based on infinite volume extra dimensions is presented.
Embedding our spacetime in infinite volume extra dimensions has
several advantages. If they are compactified, one would get a
theory approaching GR in the IR, facing Weinberg's no-go theorem
again. Details of how these large dimension models circumvent the
no-go theorem can be found in \cite{DvaliGabadadzeShifman20021}.
Moreover, often the assumption is made that the higher-dimensional
theory is supersymmetric and that susy is spontaneously broken on
the brane. These breaking effects can be localized on the brane
only, without affecting the bulk, because the infinite volume
gives a large enough suppression factor. Apart from that, an
unbroken R-parity might be assumed to forbid any negative vacuum
energy density in the bulk.

They start with the following low-energy effective action:
\begin{equation}\label{infvol}
S = M_{\ast}^{2+N}\int d^4xd^Ny\sqrt{G}\mathcal{R} + \int
d^4x\sqrt{g}\left(\mathcal{E}_4 + M_{P}^{2}R +
\mathcal{L}_{SM}\right),
\end{equation}
where $M_{\ast}^{2+N}$ is the $(4+N)$-dimensional Planck mass, the
scale of the higher dimensional theory, $G_{AB}$ the
$(4+N)$-dimensional metric, $y$ are the `perpendicular'
coordinates and $\mathcal{E}_4 = M_{Pl}^2\Lambda$, the brane
tension, or 4D cosmological constant. Thus the first term is the
bulk Einstein-Hilbert action for $(4+N)$-dimensional gravity and
the $M_{P}^{2}R$ term is the induced 4D-Einstein-Hilbert action.
So there are two free parameters: $M_{\ast}$ and $\mathcal{E}$.
$M_{\ast}$ is assumed to be very small, making gravity in the
extra dimensions much stronger than in our 4D world. The 4D-Planck
mass in this setup is a derived quantity \cite{Sakharov:1967pk}.

Gravity on the brane can be recovered either by making a
decomposition into Kaluza-Klein modes, or by considering the 4D
graviton as a resonance, a metastable state with a mass given by
$m_g\sim M_{\ast}^3/M_{Pl}^2$.

The higher dimensional graviton can be expanded in 4D Kaluza-Klein
modes as follows:
\begin{equation}
h_{\mu\nu}(x,y_n) = \int
d^Nm\epsilon_{\mu\nu}^{m}(x)\sigma_m(y_n),
\end{equation}
where $\epsilon_{\mu\nu}^{m}(x)$ are 4D spin-2 fields with mass
$m$ and $\sigma_m(y_n)$ are their wavefunction profiles in the
extra dimensions. Each of these modes gives rise to a Yukawa-type
gravitational potential, the coupling-strength to brane sources of
which are determined by the value of $\sigma_m$ at the position of
the brane, say $y=0$:
\begin{equation}
V(r)\propto\frac{1}{M_{\ast}^{2+N}}\int_{0}^{\infty}dmm^{N-1}|\sigma_m(0)|^2\frac{e^{-rm}}{r}.
\end{equation}
However, in this scenario there is a cut-off of this integral;
modes with $m>1/r_c$ have suppressed wavefunctions, where $r_c$ is
some cross-over scale, given by $r_c=M_{Pl}^2/M_{\ast}^3\sim
H_{0}^{-1}$. For $r\ll r_c$ the gravitational potential is $1/r$,
dominated by the induced 4D kinetic term, and for $r\gg r_c$ it
turns to $1/r^2$, in case of one extra dimension. In ordinary
extra dimensional gravity, all $|\sigma_m(0)|=1$, here however:
\begin{equation}
|\sigma_m(0)|=\frac{4}{4+m^2r_{c}^{2}},
\end{equation}
which decreases for $m\gg r_c$. Therefore, the gravitational
potential interpolates between the 4D and 5D regimes at $r_c$.
Below $r_c$ almost normal 4D gravity is recovered, while at larger
scales it is effectively 5-dimensional and thus weaker. This could
cause the universe's acceleration.

The question now is, whether there exist solutions such that the
4D induced metric on the brane is flat:
$g_{\mu\nu}=\eta_{\mu\nu}$. Einstein's equation from
(\ref{infvol}) now becomes (up to two derivatives):
\begin{eqnarray}
&& M_{\ast}^{2+N}\left(\mathcal{R}_{AB} -
\frac{1}{2}G_{AB}\mathcal{R}\right) + \delta^{(N)}M_{P}^2\left(R -
\frac{1}{2}g_{\mu\nu}R\right)\delta_{A}^{\mu}\delta_{B}^{\nu}\nonumber\\
&=&
\mathcal{E}_4\delta^{(N)}(y)g_{\mu\nu}\delta_{A}^{\mu}\delta_{B}^{\nu}.
\end{eqnarray}
In case of one extra dimension it is not possible to generate a
viable dynamics with a flat 4D metric. For $N\geq 2$, however,
solutions of the theory can be parameterized as:
\begin{equation}
ds^2=A^2(y)g_{\mu\nu}(x)dx^{\mu}dx^{\nu} - B^2(y)dy^2 -
C^2(y)y^2d\Omega^2_{N-1},
\end{equation}
where $y\equiv \sqrt{y_1^2 + \dots +y_n^2}$ and the functions
$A,B,C$ depend on $\mathcal{E}_4$ and $M_{\ast}$:
\begin{equation}
A,B,C=\left(1-\left(\frac{y_g}{y}\right)^{N-2}\right)^{\alpha,\beta,\gamma},
\end{equation}
where $\alpha,\beta,\gamma$ correspond to $A,B,C$ respectively,
and depend on dimensionality and $y_g$ is the gravitational radius
of the brane:
\begin{equation}
y_g\sim
M_{\ast}^{-1}\left(\frac{\mathcal{E}_4}{M_{\ast}^4}\right)^{\frac{1}{N-2}}\quad\quad\mbox{for}\quad\quad
N\neq 2.
\end{equation}
Most importantly, one explicitly known solution, with $N=2$,
generates a flat 4D Minkowski metric and $R(g)=0$
\cite{Sundrumflat98}. The 4D brane tension is spent on creating a
deficit angle in the bulk. However, one has to fine-tune this
tension in order not to generate a deficit angle larger than
$2\pi$. So also the $N=2$ model does not work.

For $N>2$ consistent solutions possibly do exist with a flat 4D
metric. However, these are not the only solutions, and besides,
their interpretation is rather complicated because of the
appearance of a naked singularity. Spacetime in $4+N$ dimensions
looks like $\Re_4\times S_{N-1}\times R_+$, where $\Re_4$ denotes
flat spacetime on the brane, and $S_{N-1}\times R_+$ are
Schwarzschild solutions in the extra dimensions.

They argue that the final physical result is:
\begin{equation}
H\sim M_{\ast}
\left(\frac{M_{\ast}^4}{\mathcal{E}_4}\right)^{\frac{1}{N-2}}.
\end{equation}
According to the 4D result, $N=0$, the expansion rate grows as
$\mathcal{E}_4$ increases, but for $N>2$ the acceleration rate $H$
decreases as $\mathcal{E}_4$ increases. In this sense, vacuum
energy can still be very large, it just gravitates very little; 4D
vacuum energy is supposed to curve mostly the extra dimensions.

This scenario has been criticized for different reasons, which we
will come to in section \ref{massgrav}. The most important issue
raised is that, since gravity has essentially become massive in
this scenario, the graviton has five degrees of freedom, and
especially the extra scalar degree of freedom, often leads to
deviations of GR at small scales.

\subsubsection{Non-local Gravity}

From a 4D-perspective, this approach can also be viewed as to make
the effective Newton's constant frequency and wavelength
dependent, in such a way that for sources that are uniform in
space and time it is tiny \cite{ArkaniHamedDimDvaGab}:
\begin{equation}\label{Ghighpass1}
M_{Pl}^2\left(1+\mathcal{F}(L^2\nabla^2)\right)G_{\mu\nu} =
T_{\mu\nu}.
\end{equation}
Here $\mathcal{F}(L^2\nabla^2)$ is a filter function:
\begin{eqnarray}
\mathcal{F}(\alpha)\rightarrow 0\quad &\mbox{for}&\; \alpha\gg
1\nonumber\\
\mathcal{F}(\alpha)\gg 1\quad &\mbox{for}&\; \alpha\ll 1
\end{eqnarray}
$L$ is a distance scale at which deviations from general
relativity are to be expected and
$\nabla^2\equiv\nabla_{\mu}\nabla^{\mu}$ denotes the covariant
d'Alembertian. Thus (\ref{Ghighpass1}) can be viewed as Einstein's
equation with $(8\pi G_{N}^{eff})^{-1} = M_{Pl}^2(1+\mathcal{F})$.
It is argued that for vacuum energy $\mathcal{F}(0)$ is large
enough, such that it will barely gravitate, resulting in a very
small curvature radius $R$:
\begin{equation}
M_{P}^2\left(1+\mathcal{F}(0)\right)G_{\mu\nu}=\left(M_{P}^2 +
\bar{M}^2\right)G_{\mu\nu},\quad\quad\mbox{and}\quad\quad R=-
\frac{4\mathcal{E}_4}{M_{P}^2 + \bar{M}^2}.
\end{equation}
To reproduce the observed acceleration a value $\bar{M}$ is needed
$\bar{M}\sim 10^{48}$~GeV for a vacuum energy density of TeV
level, and a $\bar{M}\sim 10^{80}$~GeV for $\mathcal{E}_4$ of
Planck mass value, which is about equal to the mass of the
universe.

In terms of the graviton propagator, it gets an extra factor
$(1+\mathcal{F}(k^2L^2))^{-1}$ and therefore goes to zero when
$\mathcal{F}(0)\rightarrow\infty$, instead of generating a
tadpole.

In the limit $L\rightarrow\infty$ one arrives at:
\begin{equation}
M_{Pl}^2G_{\mu\nu} - \frac{1}{4}\bar{M}^2g_{\mu\nu}\bar{R} =
T_{\mu\nu},
\end{equation}
just the zero mode part of $G_{\mu\nu}$, which is proportional to
$g_{\mu\nu}$, where
\begin{equation}
\bar{R}\equiv\frac{\int d^4x\sqrt{g}R}{\int d^4x\sqrt{g}}
\end{equation}
$\bar{R}$ thus is the spacetime averaged Ricci curvature, which
vanishes for all localized solutions, such as stars, black holes
and also for FRW models. For de Sitter space however, $\bar{R}\neq
0$., but a constant and equal to $\bar{R}=R_{\infty}$, with
$R_\infty$ the asymptotic de Sitter curvature.

At the price of losing 4D-locality and causality, the new averaged
term is both non-local and acausal, a model is constructed in
which a huge vacuum energy does not lead to an unacceptably large
curvature. The Planck scale is made enormous for Fourier modes
with a wavelength larger than a size $L$. Such sources would feel
gravity only due to their coupling with the graviton zero mode.
This zero mode however, is very weakly coupled to brane sources
since it is suppressed by the volume of the extra dimensions.

It is argued that the acausality has no other observable effect.
Moreover, it has been claimed that non-locality should be an
essential element in any modification of GR in the infrared that
intends to solve the cosmological constant problem
\cite{Arkani-HamedDimopoulosDvaliGabadadze}. The argument is that
it takes local, causal physics a time $1/L$ to respond to
modifications at scale $L\sim 10^{28}$~cm, and thus in particular
to sources which have characteristic wavelength larger than
$H_{0}^{-1}$, ``such as vacuum energy"
\cite{DvaliGabadadzeShifman20022}.

The non-localities in this case appear in the four dimensional
truncation of the $4+N$-dimensional theory of section
(\ref{infiniteed}). There is an infinite number of degrees of
freedom below any non-zero energy scale. Therefore, in order to
rewrite the model as an effective four dimensional field theory,
and infinite number of degrees of freedom have to be integrated
out. This results in the appearance of non-local interactions,
despite the fact that the full theory is local.

Another idea based on a model of non-local quantum gravity and
field theory due to Moffat \cite{Moffat,Moffat1}, also suppresses
the coupling of gravity to the vacuum energy density and also
leads to a violation of the Weak Equivalence Principle.

\subsubsection{Massive Gravitons}\label{massgrav}

A much studied approach to change general relativity in the
infrared which is not simply a variety of a scalar-tensor theory,
is to allow for tiny masses for gravitons, like in the Fierz-Pauli
theory of massive gravity \cite{FierzPauli1939}, and in the
example above. Note in passing that due to mass terms, gravitons
might become unstable and could possibly decay into lighter
particles, for example photons. If so, gravity no longer obeys the
standard inverse-square law, but becomes weaker at large scales,
leading to accelerated cosmic expansion.

Of course, the extra degrees of freedom, extra polarizations of a
massive graviton, could also become noticeable at much shorter
distances, putting severe constraints on such scenarios. In the UV
the new scalar degrees of freedom become strongly coupled, where
the effective theory breaks down and the physics becomes sensitive
to the unknown UV-completion of the theory.

A severe obstacle massive gravity theories have to overcome is
something known as the Van Dam, Veltman, Zakharov, or (vDVZ),
discontinuity \cite{vanDam:1970vg,Zakharov}. vDVZ argued that in
the massive case, even with extremely small graviton mass, the
bending of light rays passing near the sun would be too far off
from experimental results in the massive case, that the mass of
the graviton has to be exactly equal to zero. The physical reason
indeed being, that even in the limit where the mass of the
graviton goes to zero, there is an additional scalar attraction,
which distinguishes the theory from Einstein's GR.

In the DGP model, the extra dimensions are infinitely large, and
in the literature, there is an ongoing discussion whether this
model is experimentally viable and capable of avoiding the massive
gravity difficulties, see
\cite{Luty:2003vm,Rubakov:2003zb,Porrati:2004yi,Nicolis:2004qq,Deffayet:2005ys,Lue:2005ya}
for criticism. It appears that indeed also this model suffers from
strong interactions at short distances due to the scalar
polarization of the massive graviton, that can be understood in
terms of a propagating ghosts-like degree of freedom.

The deviations of GR are argued to take place at distances set by
$r_c\equiv M_{P}^2/M_{\ast}^3$. The one-graviton exchange
approximation breaks down at distances $R_{\ast}\sim (R_s
r_{c}^2)^{1/3}$ \cite{Luty:2003vm}, called the Vainshtein scale,
with $R_S$ the Schwarzschild radius of the source. $R_{\ast}$ is
very large for astrophysical sources, which suggests that the DGP
model may describe our universe. For distances larger than
$R_{\ast}$ gravity deviates significantly from GR, yet for smaller
distances it should yield (approximately) the same results.
However, quantum effects become important at much smaller
distances scales, given by:
\begin{equation}
r_{crit}=\left(\frac{r_{c}^2}{M_{P}^2}\right)^{1/3},
\end{equation}
which can be as small as a 1000~km, for $r_c\sim H\sim
10^{28}$~cm. These strong interactions can be traced back to the
appearance of a negative norm state. This is however a
controversial result, also argued for in \cite{Rubakov:2003zb},
yet waived away in \cite{Koyama:2005tx}. Further studies are
necessary to settle this question.

The Schwarzschild solutions in the DGP model are also heavily
debated and it is not yet clear what the correct way is to
calculate these, and whether they will eventually lead to
consistent phenomenological behavior. For a recent study and
references, see \cite{Gabadadze:2005qy}.

In the next section we will consider an alternative, that does not
suffer from this `strong coupling problem'.

In \cite{Will:2004xi} bounds on graviton masses are discussed
using the LISA space interferometer.

\subsection{Ghost Condensation or Gravitational Higgs Mechanism }

In this framework gravity is modified in the infrared as a result
of interactions with a `ghost condensate', leading among other
things to a mass for the graviton, see \cite{Arkani-Hamed-ghost}.

Assume that for a scalar field $\phi$ we have:
\begin{equation}
\langle\dot{\phi}\rangle = M^2,\quad\rightarrow\quad\phi = M^2t +
\pi
\end{equation}
and that it has a shift symmetry $\phi\rightarrow\phi + a$ so that
it is derivatively coupled, and that its kinetic term enters with
the wrong sign in the Lagrangian:
\begin{equation}
\mathcal{L}_{\phi} =
-\frac{1}{2}\partial^{\mu}\phi\partial_{\mu}\phi + \dots
\end{equation}
The consequence of this wrong sign is that the usual background
with $\langle\phi\rangle = 0$ is unstable and that after vacuum
decay, the resulting background will break Lorentz invariance
spontaneously.

The low energy effective action for the $\pi$ has the form:
\begin{equation}
S\sim\int
d^4x\left[\frac{1}{2}\dot{\pi}^2-\frac{1}{2M^2}(\nabla^2\pi)^2
+\dots\right],
\end{equation}
so that the $\pi$'s have a low energy dispersion relation like:
\begin{equation}\label{ghostdispers}
\omega^2\sim\frac{k^4}{M^2}
\end{equation}
instead of the ordinary $\omega^2\sim k^2$ relation for light
excitations. Time-translational invariance is broken, because
$\langle\phi\rangle = M^2t$ and as a consequence there are two
types of energy, a ``particle physics" and a ``gravitational"
energy which are not the same. The particle physics energy takes
the form:
\begin{equation}
\mathcal{E}_{pp}\sim\frac{1}{2}\dot{\pi}^2 +
\frac{\left(\nabla^2\pi\right)^2}{2M^2} +\dots,
\end{equation}
whereas the gravitational energy is:
\begin{equation}
\mathcal{E}_{grav}= T_{00}\sim M^2\dot{\pi} +\dots
\end{equation}
Although time-translation- and shift-symmetry are broken in the
background, a diagonal combination is left unbroken and generates
new ``time" translations. The Noether charge associated with this
unbroken symmetry is the conserved particle physics energy. The
energy that couples to gravity is associated with the broken time
translation symmetry. Since this energy begins at linear order in
$\dot{\pi}$, lumps of $\pi$ can either gravitate or
anti-gravitate, depending on the sign of $\dot{\pi}$! The $\pi$
thus maximally violate the equivalence principle.

If the standard model fields would couple directly to the
condensate there would be a splitting between particle and
anti-particle dispersion relations, and a new spin-dependent
inverse-square force, mediated by $\pi$ exchange, which results
from the dispersion relation (\ref{ghostdispers}). In the
non-relativistic limit, with $\mathbf{S}$ the spin:
\begin{equation}
\triangle\mathcal{L} \sim\frac{1}{F}\mathbf{S}\cdot\nabla\pi,
\end{equation}
where $F$ is some normalization constant. Because of the $k^4$
dispersion relation, the potential between two sources with spin
$\mathbf{S}_1$ and spin $\mathbf{S}_2$, will be proportional to
$1/r$:
\begin{equation}
V\sim\frac{M^4}{\tilde{M}^2F^2}\frac{\mathbf{S}_1\cdot\mathbf{S}_2-3\left(\mathbf{S}_1\cdot\hat{\mathbf{r}}\right)}{r},
\end{equation}
when using only static sources, ignoring retardation effects.

Moreover, not only Lorentz invariance, but also CPT is broken if
the standard model fields would couple directly to the condensate.
The leading derivative coupling is of the form:
\begin{equation}
\triangle\mathcal{L} =
\sum_{\psi}\frac{c_{\psi}}{F}\bar{\psi}\bar{\sigma}^{\mu}\psi\partial_{\mu}\phi.
\end{equation}
As noted in \cite{Arkani-Hamed-ghost}, field redefinitions
$\psi\rightarrow e^{ic_{\psi}\phi/F}\psi$ may remove these
couplings, but only if such a $U(1)$ symmetry is not broken by
mass terms or other couplings in the Lagrangian. If the fermion
field $\psi$ has a Dirac mass term $m_D\psi\psi^c$, then the
vector couplings, for which $c_{\psi} + c_{\psi^c} =0$, still can
be removed, but the axial couplings remain:
\begin{equation}
\triangle\mathcal{L}\sim\frac{1}{F}\bar{\Psi}\gamma^{\mu}\gamma^5\Psi\partial_{\mu}\phi.
\end{equation}
After expanding $\phi = M^2t +\pi$ this becomes:
\begin{equation}
\triangle\mathcal{L}\sim \mu\bar{\Psi}\gamma^0\gamma^5\Psi +
\frac{1}{F}\bar{\Psi}\gamma^{\mu}\gamma^5\Psi\partial_{\mu}\pi,
\end{equation}
with $\mu = M^2/F$. This first term violates both Lorentz
invariance and CPT, leading to different dispersion relations for
particles and their anti-particles. A bound on $\mu$ is obtained
by considering the earth to be moving with respect to spatially
isotropic condensate background. The induced Lorentz and CPT
violating mass term then looks like:
\begin{equation}
\mu\bar{\Psi}\mathbf{\gamma}\gamma^5\Psi\cdot\mathbf{v}_{earth},
\end{equation}
which in the non-relativistic limit gives rise to an interaction
Hamiltonian:
\begin{equation}
\mu\mathbf{S}\cdot\mathbf{v}_{earth}.
\end{equation}
The experimental limit on $\mu$ for coupling to electrons is
$\mu\leq 10^{-25}$~GeV \cite{Heckeletal1999} assuming
$|\mathbf{v}_{earth}|\sim 10^{-3}$. For other limits on CPT and
Lorentz invariance, see
\cite{Phillipsetal2000,Bluhm2003,Cane2003}.

If there is no direct coupling, the SM fields would still interact
with the ghost sector through gravity. Interestingly, IR
modifications of general relativity could be seen at relatively
short distances, but only after a certain (long) period of time!
Depending on the mass $M$ and the expectation value of $\phi$,
deviations of Newtonian gravity could be seen at distances
1000~km, but only after a time $t_c\sim H_{0}^{-1}$ where $H_0$ is
the Hubble constant. More general, the distance scale at which
deviations from the Newtonian potential are predicted is $r_c\sim
M_{Pl}/M^2$ and their time scale is $t_c\sim M_{Pl}^2/M^3$.

To see the IR modifications to GR explicitly, let us consider the
effective gravitational potential felt by a test mass outside a
source $\rho_m(r,t)=\delta^3(r)\theta(t)$, i.e. a source that
turns on at time $t=0$. This potential is given by:
\begin{equation}
\Phi(r,t)=-\frac{G}{r}\left[1+I(r,t)\right],
\end{equation}
where $I(r,t)$ is a spatial Fourier integral over momenta $k$,
evaluated using an expansion around flat space; a bare
cosmological constant is set to zero.
\begin{eqnarray}
I(r,t)=\frac{2}{\pi}&\{&\int_{0}^{1}du\frac{\sin(uR)}{(u^3-u)}\left(1-\cosh(Tu\sqrt{1-u^2})\right)\nonumber\\
&+&\int_{1}^{\infty}du\frac{\sin(uR)}{(u^3-u)}\left(1-\cos(Tu\sqrt{u^2-1})\right)\}.
\end{eqnarray}
Here $u=k/m$, $R=mr$, $T=\alpha M^3/2M_{Pl}^2$, where $m\equiv
M^2/\sqrt{2}M_{Pl}$ and $\alpha$ is a coefficient of order 1. For
late times, $t\gtrsim t_c$, or $T\gtrsim 1$, the first integrand
will dominate and $I(r,t)$ can be well approximated by:
\begin{equation}
I(r,t) \simeq \frac{2}{\sqrt{\pi
T}}\exp\left(-\frac{R^2}{8T}+\frac{T}{2}\right)\sin\left(\frac{R}{\sqrt{2}}\right).
\end{equation}
For $R\ll T$, there is indeed an oscillatory behavior for the
gravitational potential, growing exponentially as $\exp(T/2)$,
while for $R\gg T$ the modification vanishes.

More general gravitational effects have been studied in
\cite{Dubovsky2004}, where moving sources were considered, and in
\cite{Arkani-Hamed2003} where inflation was studied in this
context. Moreover, the quantum stability of the condensate was
studied in \cite{Krotovetal2004}.

This highly speculative scenario opens up a new way of looking at
the cosmological constant problem, especially because of the
distinction between particle physics energy, $\mathcal{E}_{pp}$
and gravitational energy, $\mathcal{E}_{grav}$. It has to be
developed further to obtain a better judgement.

\subsection{Fat Gravitons}\label{fatgravitons}

\begin{figure}[h]
\includegraphics[width=15cm]{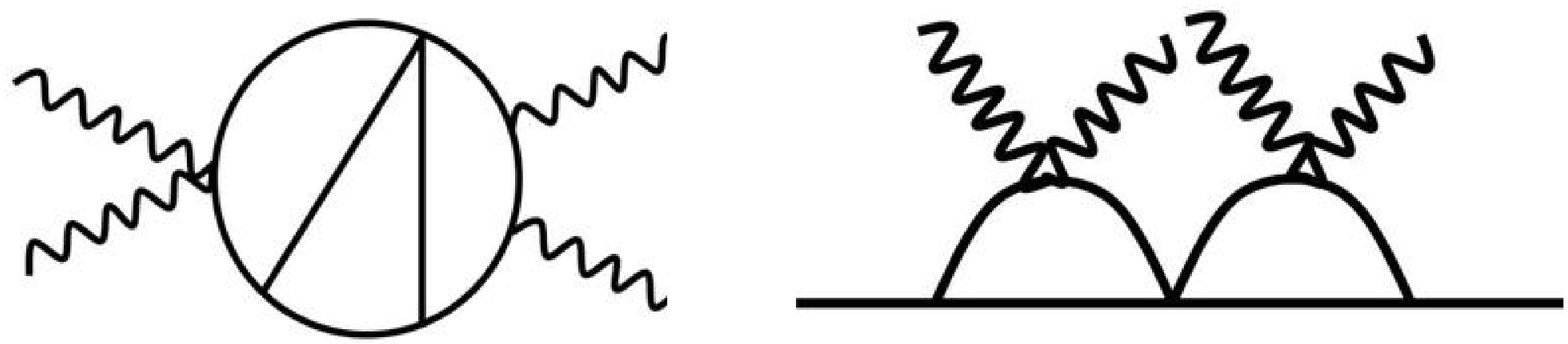}
\caption{On the left-hand-side, a typical Standard Model
contribution to $\Gamma_{eff}[g_{\mu\nu}]$. On the right, soft
gravitons coupled to loop-correction to SM self-energy. Wiggly
lines are gravitons and smooth lines are SM particles.}
\label{gravselfen}
\end{figure}

A proposal involving a sub-millimeter breakdown of the
point-particle approximation for gravitons has been put forward by
Sundrum \cite{Sundrum}. In standard perturbative gravity, diagrams
with external gravitons and SM-particles in loops (see figure
\ref{gravselfen})
give a contribution to the effective CC of which the dominant part
diverges as $\Lambda^4_{UV}$ where $\Lambda_{UV}$ is some
ultraviolet cutoff. This leads to the enormous discrepancy with
experimental results for any reasonable value of $\Lambda_{UV}$.
However, one might wonder what the risks are when throwing away
these diagrams from the effective theory
$\Gamma_{eff}[g_{\mu\nu}]$, when $|k^2|$, the momentum of the
external gravitons, is larger than some low energy cutoff.
Properties at stake are: Unitarity, General Coordinate Invariance
(GCI) and locality. In standard effective theory one also has
diagrams where soft gravitons give corrections to the SM self
energy diagrams, (figure \ref{gravselfen}). These cannot be thrown
away, since they are crucial in maintaining the equivalence
principle between inertial and gravitational masses. However,
locally these diagrams are indistinguishable in spacetime, only
globally can we discern their topological difference. Thus given
locality of the couplings of the point particles in the diagrams,
we cannot throw the first diagram away and keep the other.
Therefore, it seems progress can be made by considering a graviton
as an extended object. Define the graviton size:
\begin{equation}
l_{grav}\equiv \frac{1}{\Lambda_{grav}}.
\end{equation}
Such a ``fat graviton" does not have to couple with point-like
locality to SM loops, but with locality up to $l_{grav}$. Thus a
fat graviton can distinguish between the two types of diagrams,
possibly suppressing the first while retaining the second.

The value of the CC based on usual power counting would then be:
\begin{equation}
\Lambda_{eff}\sim\mathcal{O}(\Lambda^{4}_{grav}/16\pi^{2}).
\end{equation}
Comparing with the observational value this gives a bound on the
graviton size of:
\begin{equation}
l_{grav} > 20~\mbox{microns}
\end{equation}
which would indicate a short-distance modification of Newton's law
below $20$~microns. This is however not enough to suppress
standard model contributions to the cosmological constant. A new
model by the same author has been proposed to take into account
also these effects, see section \ref{Energytominus}.

\subsection{Composite Graviton as Goldstone boson}

Another approach is to consider the possibility that the graviton
appears as a composite Goldstone boson. There exists a theorem by
Weinberg and Witten, \cite{WeinbergWitten1980}, stating that a
Lorentz invariant theory, with a Lorentz covariant energy-momentum
tensor does not admit a composite graviton. It is therefore
natural to try a mechanism where the graviton appears as a
Goldstone boson associated with the spontaneous breaking of
Lorentz invariance. Being a Goldstone boson, the graviton would
not develop a potential, and hence the normal cosmological
constant problem is absent, see for example \cite{KrausTomboulis,
Jenkins}.

However, besides difficulties erasing the traces of broken Lorentz
invariance to make the model agree with observations, also new
fine-tunings are introduced.

A composite structure of the graviton has also been contemplated
in \cite{Zee:2003dz,Zee:2004um}, based on more intuitive ideas.

\subsection{Summary}

Since General Relativity has only been thoroughly tested on solar
system distance scales it is a very legitimate idea to consider
corrections to GR at galactic and/or cosmological distance scales.
However, often these models are not so harmless as supposed to be.
The laws of gravity are also significantly changed at shorter
scales, or the changes lead to violations of locality. The
scenarios described in this section do not directly solve the
cosmological constant problem, but offer new ways of looking at
it.

On the more positive side, many theories that predict
modifications of GR in the IR, reproduce Einstein gravity at
smaller distances, but up to some small corrections. These
corrections are discussed in \cite{DvaliGruzinovZaldarriagaMoon}
and could be potentially observable at solar system distance
scales. At the linearized level gravity is of the scalar-tensor
type, because the graviton has an extra polarization that also
couples to conserved energy-momentum sources. If these models are
correct, an anomalous perihelion precession of the planets is
expected to be observed in the near future.

Besides, submillimeter experiments of Newtonian gravity set ever
more stringent bounds on both extra dimensional approaches and
composite graviton scenarios. It would be very exciting to see a
deviation of Newtonian gravity at short distances. On the other
hand, observing no change at all, will seriously discourage the
hopes that such a mechanism might help in solving the cosmological
constant problem.

\section{Type IV: Statistical Approaches}

\subsection{Hawking Statistics}\label{Hawking}

If the cosmological constant could a priori have any value,
appearing for example as a constant of integration as in section
(\ref{unimodular}), or would become a dynamical variable by means
of some other mechanism, then in quantum cosmology the state
vector of the universe would be a superposition of states with
different values of $\Lambda_{eff}$. The path integral would
include all, or some range of values of this effective
cosmological constant. The observable value of the CC in this
framework is not a fundamental parameter. Different universes with
different values of $\Lambda_{eff}$ contribute to the path
integral. The probability of observing a given field configuration
will be proportional to $\mathcal{P}\propto\exp
(-S(\Lambda_{eff}))$ in which $\Lambda_{eff}$ is promoted to be a
quantum number.

Eleven dimensional supergravity contains a three-form gauge field,
with a four-form field strength $F_{\mu\nu\rho\sigma} =
\partial_{[\mu}A_{\nu\rho\sigma ]}$ \cite{Aurilia:1980xj}. When
reduced to four dimensions, this gives a contribution to the
cosmological constant
\cite{Duff:1980qv,Baum,GomberoffHenneauxTeitelboimWilczek,BrownTeitelboim1,BrownTeitelboim2}.
Hawking \cite{Hawking1} used such a three-form gauge field to
argue that the wave function of the universe is peaked at zero
cosmological constant. It is the first appearance of the idea that
the CC could be fixed by the shape of the wave function of the
universe.

The three-form field $A_{\mu\nu\lambda}$ has gauge
transformations:
\begin{equation}
A_{\mu\nu\rho}\rightarrow A_{\mu\nu\rho}
+\nabla_{[\mu}C_{\nu\rho]},\quad\quad\mbox{with}\quad\quad
F_{\mu\nu\rho\sigma}=\nabla_{[\mu}A_{\nu\rho\sigma]}.
\end{equation}
This field would contribute an extra term to the action:
\begin{equation}
I = -\frac{1}{16\pi G}\int d^4x\sqrt{-g}\left( R+2\Lambda_B\right)
-\frac{1}{48}\int
d^4x\sqrt{-g}F_{\mu\nu\rho\sigma}F^{\mu\nu\rho\sigma}.
\end{equation}
The field equation for $F^{\mu\nu\rho\sigma}$ is:
\begin{equation}\label{solHawk}
D_{\mu}F^{\mu\nu\rho\sigma}=0,\quad\quad\rightarrow\quad\quad
\sqrt{-g}F^{\mu\nu\rho\sigma} = \omega\epsilon^{\mu\nu\rho\sigma}
\end{equation}
Such a field $F$ has no dynamics, but the $F^2$ term in the action
behaves like an effective cosmological constant term, whose value
is determined by the unknown scalar field $\omega$, which takes on
some arbitrary value. If we substitute the solution
(\ref{solHawk}) back into the Einstein equation, we find, using
that $\epsilon^{\mu\nu\rho\sigma}\epsilon_{\mu\nu\rho\sigma}=\pm
4!$:
\begin{equation}
T^{\mu\nu}= \frac{1}{6}\left(
F^{\mu\alpha\beta\gamma}F^{\nu}_{\alpha\beta\gamma} -
\frac{1}{8}g^{\mu\nu}F^{\alpha\beta\gamma\delta}F_{\alpha\beta\gamma\delta}\right)
= \pm \frac{1}{2}\omega^2g^{\mu\nu}
\end{equation}
where the sign depends on the metric used: in Euclidean metric
$\epsilon^{\mu\nu\rho\sigma}\epsilon_{\mu\nu\rho\sigma}$ is
positive, whereas in Lorentzian metric it is negative. In the
Euclidean action Hawking used:
\begin{equation}
R=-4\Lambda_{eff}=-4(\Lambda_B - 8\pi G\omega^2)
\end{equation}
where $\Lambda_B$ is the bare cosmological constant in Einstein's
equation. It follows that:
\begin{equation}\label{effstat}
S_{Hawking} = -\Lambda_{eff}\frac{V}{8\pi G}.
\end{equation}
The maximum value of this action is given when $V$ is at its
maximum, which Hawking takes to be $S^4$, with radius
$r=(3\Lambda^{-1}_{eff})^{1/2}$ and proper circumference $2\pi r$.
Then:
\begin{equation}
V=\frac{24\pi^2}{\Lambda_{eff}^2},\quad\quad\rightarrow\quad\quad
S(\Lambda)=-3\pi\frac{M^{2}_{P}}{\Lambda_{eff}}
\end{equation}
and thus the probability density:
\begin{equation}
\mathcal{P}\propto
\exp\left(3\pi\frac{M^{2}_{P}}{\Lambda_{eff}}\right)
\end{equation}
is peaked at $\Lambda=0$.

Note that we have used here that the probability is evaluated as
the exponential of minus the effective action at its stationary
point. That is, stationary in $A_{\mu\nu\lambda}$, meaning
vanishing covariant derivative of $F_{\mu\nu\lambda\rho}$, in
matter fields $\phi$ and in $g_{\mu\nu}$. The latter condition
simply means that $g_{\mu\nu}$ has to satisfy the Einstein
equations. Eqn. (\ref{effstat}) is the effective action at the
stationary point. It is a good thing that we only need the
effective action at its stationary point, so that we do not have
to worry about the Euclidean action not being bounded from below,
see for example \cite{Weinbergreview}.

However, Hawking's argument is not correct, since one should not
plug an ans\"{a}tz for a solution back into the action, but rather
vary the unconstrained action \cite{Duff:1989ah}. This differs a
minus sign in this case, the same minus sign as going from a
Lorentzian to a Euclidean metric, $\Lambda_{eff}=(\Lambda_B \pm
8\pi G\omega^2)$, but now between the coefficient of $g^{\mu\nu}$
in the Einstein equations, and the coefficient of $(8\pi
G)^{-1}\sqrt{g}$ in the action. The correct action becomes
\cite{Duff:1989ah}:
\begin{equation}
S =(-3\Lambda_{eff} + 2\Lambda_B)\frac{-3\pi
M_{P}^2}{\Lambda^{2}_{eff}} = -3\pi M_{P}^2\frac{\Lambda_B -12\pi
G\omega^2}{(\Lambda_B - 4\pi G\omega^2)^2}
\end{equation}\
now for $\Lambda_{eff}\rightarrow 0$, the action becomes large and
positive and consequently, $\Lambda_{eff}=0$ becomes the
\textit{least} probable configuration.

Besides, in \cite{PolchinskiBousso} it is shown that this approach
has also other serious limitations. It is argued that it can only
work in the `Landscape' scenario that we discuss in section
(\ref{AP}). The reason is that the four-form flux should be
subject to Dirac quantization and the spacing in $\Lambda$ then
only becomes small enough with an enormous number of vacua.

\subsection{Wormholes}

In a somewhat similar approach Coleman \cite{colemanworm} argued
that one did not need to introduce a 3-form gauge field, if one
includes the topological effects of wormholes. This also
transforms the cosmological constant into a dynamical variable.
The argument is that on extremely small scales our universe is in
contact, through wormholes, with other universes, otherwise
disconnected, but governed by the same physics as ours. Although
the two ends of a wormhole may be very far apart, in the effective
theory of just our universe, the only effect of wormholes is to
add \textit{local} interactions, one for each type of wormhole.

The extra term in the action has the form:
\begin{equation}\label{wormholact}
S_{wormohole} = -\sum_i(a_i+a^{\dag}_{i})\int
d^4x\sqrt{g}e^{-S_i}K_i
\end{equation}
where $a_i$ and $a^{\dag}_{i}$ are the annihilation and creation
operators for a type $i$ baby universe, $S_i$ is the action of a
semi-wormhole (one that terminates on a baby universe), and $K_i$
is some function of fields on the manifold, with an important
exponential factor that suppresses the effects of all wormholes,
except those of Planckian size
\cite{Hawking:1987mz,Strominger:1983ns,Coleman:1988cy}.

The coefficients of these interaction terms are operators $A_i=
a_i+a_{i}^{\dag}$ which only act on the variables describing the
baby universes, and commute with everything else. Written in terms
of $A$-eigenstates, the effective action becomes:
\begin{equation}\label{wormholacteigen}
S_{wormohole} = -\sum_i\int d^4x\sqrt{g}\alpha_ie^{-S_i}K_i,
\end{equation}
with $\alpha_i$ the eigenvalues of the operators $A$, which would
be interpreted as constants of nature, by an observer doing
experiments at distance scales larger than the wormhole scale,
i.e. for an observer who cannot detect the baby universes.

This way, the effective cosmological constant becomes a function
of the $a_i$. Moreover, on scales larger than the wormhole scale,
manifolds that appear disconnected will really be connected by
wormholes, and therefore are to be integrated over.

The sum of all vacuum-to-vacuum graphs is the exponential of the
sum of connected graphs, which gives the probability density
$\mathcal{P}$:
\begin{equation}\label{sumeffact}
\mathcal{P}\propto\exp\left[\sum_{CCM}e^{-S_{eff}(\alpha)}\right],
\end{equation}
where $CCM$ stands for closed connected manifolds. The sum can be
expressed as a background gravitational field effective action,
$\Gamma$. The sum over closed connected manifolds can then be
written as a sum over topologies:
\begin{equation}
\sum_{CCM}e^{-S_{eff}(\alpha)} = \sum_{topologies}e^{-\Gamma(g)},
\end{equation}
with $g$ the background metric on each topology and each term on
the right is again to be evaluated at its stationary point. This
is progress, since the leading term in $\Gamma$ for large, smooth
universes is known, and is the cosmological constant term:
\begin{equation}\label{gammaleading}
\Gamma = \Lambda(\alpha)\int d^4x\sqrt{g} + \ldots,
\end{equation}
$\Lambda(\alpha)$ being the fully renormalized cosmological
constant. Plugging this back into (\ref{sumeffact}) gives the
final result:
\begin{equation}
\mathcal{P}\propto\exp\left[\exp\left(3\pi\frac{M_{P}^2}{\Lambda_{eff}}\right)\right],
\end{equation}
and thus is even sharper peaked at $\Lambda =0$ than in Hawking's
case. For positive CC the maximum volume is taken, like in
Hawking's case, the 4-sphere with $r=(3\Lambda^{-1}_{eff})^{1/2}$.
Furthermore, the higher order terms in (\ref{gammaleading}) are
neglected.

An advantage of Coleman's approach is that he is able to sidestep
many technical difficulties Hawking's approach suffers from. In
particular, he uses the Euclidean path integral, which is a
solution to the Wheeler-DeWitt equation, only to calculate
expectation values of some scalar field. These are independent of
$x$, because the theory is generally covariant. It includes an
average over the time in the history of the universe that the
expectation value for this operator was measured. This circumvents
many issues related to the notion of time in quantum gravity.

However, both Hawking's and Coleman's proposal rely strongly on
using a Euclidean path integral and it is unclear whether this is
suitable for a theory of quantum gravity.

There is also a more direct problem with Coleman's idea, as put
forward by Fishler, Susskind and Polchinski
\cite{Fischler:1988ia,Polchinski:1989ae}, also see
\cite{Preskill:1988na,Coleman:1989ky}. The problem is that in
Coleman's scenario wormholes of every size will materialize in the
vacuum with maximum kinematically allowed density, leading to a
universe packed with wormholes of every size. The exponential
suppression factor in (\ref{wormholact}) is inconsistent with the
other assumptions that quantum gravity is described by a Euclidean
path integral, which is dominated by large scale spherical
universes connected by wormholes, where the amplitude of a large
scale universe is of order $\exp(M^{2}_{P}/\Lambda)$. In
particular, taking into account the higher order terms in
(\ref{gammaleading}), leads to a violation of the dilute gas
approximation, used by Coleman.

In conclusion, wormholes should be left out of the functional
integral of quantum gravity. Rather, their effect is that they
renormalize the values of physical constants in our universe. Most
importantly, if for some reason it is valid to only take
Planck-scale wormholes into account, this would make the
wavefunction of the universe in the Euclidean formalism, peak at
zero value of the cosmological constant.

\subsection{Anthropic Principle}\label{AP}

One of the first to use anthropic arguments related to the value
of the cosmological constant was Weinberg \cite{Weinberg:1987dv},
see also \cite{Banks:1984cw,BarrowTipler}. He even made the
prediction in 1987 that, since the anthropic bound was just a few
orders of magnitude larger than the experimental bounds, a
non-zero cosmological constant would soon be discovered, which
indeed happened.

One can rather easily set anthropic bounds on the value of the
cosmological constant. A large positive CC would very early in the
evolution of the universe lead to an exponentially expanding de
Sitter phase, which then lasts forever. If this would happen
before the time of formation of galaxies, at redshift $z\sim 4$,
clumps of matter would not become gravitationally bound, and
galaxies, and presumably intelligent life, would not form.
Therefore:
\begin{equation}
\Omega_{\Lambda}(z_{gal})\leq\Omega_M(z_{gal})\quad\quad\rightarrow\quad\quad\frac{\Omega_{\Lambda
0}}{\Omega_{M0}}\leq a_{gal}^3=(1+z_{gal})^3\sim 125.
\end{equation}
This implies that the cosmological constant could have been larger
than observed and still not be in conflict with galaxy formation
(note that in these estimates everything is held fixed, except
$\Omega_{\Lambda}$ which is allowed to vary, unless stated
otherwise).

A typical observer therefore would measure
$\rho_\Lambda\sim\tilde{\rho}_\Lambda$, with
$\tilde{\rho}_\Lambda$ the value for which the vacuum energy
density dominates at about the epoch of galaxy formation. This is
the anthropic prediction and it peaks at $\Omega_\Lambda\sim 0.9$,
in agreement with the experimental value $\Omega_\Lambda\sim 0.7$
at the $2\sigma$ level \cite{Garriga:2003hj}. It is argued that
the agreement can be increased to the $1\sigma$ level, by allowing
for non-zero neutrino masses \cite{Pogosian:2004hd}. Neutrino
masses would slow down the growth of density fluctuations, and
hence influence the value of $\tilde{\rho}_\Lambda$. The sum of
the neutrino masses would have to be $m_\nu\sim 1-2$~eV.

On the other hand, a large negative cosmological constant would
lead to a rapid collapse of the universe and (perhaps) a big
crunch. To set this lower anthropic bound, one has to wonder how
long it takes for the emergence of intelligent life. If 7~billion
years is sufficient, the bound for a flat universe is
$\Lambda\gtrsim -18.8~\rho_0\sim -2\times 10^{-28}~\mbox{g/cm}^3$,
if 14~billion years are needed, the constraint is $\Lambda\gtrsim
-4.7~\rho_0\sim -5\times 10^{-29}~\mbox{g/cm}^3$
\cite{Kallosh:2002gg}.

It makes more sense however, to ask what the most likely value of
the cosmological constant is, the value that would be experienced
by the largest number of observers. Vilenkin's ``Principle of
Mediocrity'' \cite{Vilenkin2}, stating that we should expect to
find ourselves in a big bang that is typical of those in which
intelligent life is possible, is often used. In order for such
statistics to be meaningful, it is necessary that there are
alternative conditions where things are different. Therefore, it
is usually assumed that there is some process that produces an
ensemble of a large number of universes, or different, isolated
pockets of the same universe, with widely varying properties.
Several inflationary scenarios
\cite{Linde3,Linde4,LindeMezhlumian,Linde:1986fc}, quantum
cosmologies, \cite{Hawking1, Coleman2, Linde5, Vilenkin2,
Garcia-BellidoLinde} and string theory
\cite{PolchinskiBousso,Susskind1,Susskind2,Kachru:2003aw,Denef:2004ze,FreivogelSusskind}
predict different domains of the universe, or even different
universes, with widely varying values for the different coupling
constants. In these considerations it is assumed that there exists
many discrete vacua with densely spaced vacuum energies.

The probability measure for observing a value $\rho_\Lambda$,
using Bayesian statistics, can be written as:
\begin{equation}\label{BayesCC}
d\mathcal{P}(\rho_\Lambda) =
N(\rho_\Lambda)\mathcal{P}_{\ast}(\rho_{\Lambda})d\rho_\Lambda,
\end{equation}
where $\mathcal{P}_{\ast}(\rho_{\Lambda})d\rho_\Lambda$ is the a
priori probability of a particular big bang having vacuum energy
density between $\rho_\Lambda$ and $\rho_\Lambda + d\rho_\Lambda$
and is proportional to the volume of those parts of the universe
where $\rho_\Lambda$ takes values in the interval $d\rho_\Lambda$.
$N(\rho_\Lambda)$ is the average number of galaxies that form at a
specified $\rho_\Lambda$ \cite{Carroll2000}, or, the average
number of scientific civilizations in big bangs with energy
density $\rho_\Lambda$ \cite{Weinberg2000}, per unit volume. The
quantity $N(\rho_\Lambda)$ is often assumed to be proportional to
the number of baryons, that end up in galaxies.

Given a particle physics model which allows $\rho_\Lambda$ to
vary, and a model of inflation, one can in principle calculate
$\mathcal{P}_{\ast}(\rho_{\Lambda})$, see the above references for
specific models and \cite{Vilenkin:2004fj} for more general
arguments. $\mathcal{P}_{\ast}(\rho_{\Lambda})d\rho_\Lambda$ is
sometimes argued to be constant \cite{Martel:1997vi}, since
$N(\rho_\Lambda)$ is only non-zero for a narrow range of values of
$\rho_\Lambda$. Others point out that there may be a significant
departure from a constant distribution \cite{Garriga:1999bf}. Its
value is fixed by the requirement that the total probability
should be one:
\begin{equation}
d\mathcal{P}(\rho_\Lambda) =
\frac{N(\rho_\Lambda)d\rho_\Lambda}{\int
N(\rho'_\Lambda)d\rho'_\Lambda}.
\end{equation}
The number $N(\rho_\Lambda)$ is usually calculated using the
so-called `spherical infall' model of Gunn and Gott
\cite{Gunn:1972sv}. Assuming a constant
$\mathcal{P}_{\ast}(\rho_{\Lambda})$, it is argued that the
probability of a big bang with $\Omega_\Lambda\lesssim 0.7$ is
roughly 10\%, depending on some assumptions about the density of
baryons at recombination \cite{Weinberg2000,Weinberg:2005fh}.

However, it has been claimed that these successful predictions
would not hold, when other parameters, such as the amplitude of
primordial density fluctuations are also allowed to vary
\cite{Banks:2003es,Graesser:2004ng}. These arguments are widely
debated and no consensus has been reached
\cite{Garriga:2005ee,Feldstein:2005bm}.

However, it has been very difficult to calculate the a priori
distribution. The dynamics, leading to a ``multiverse'' in which
there are different pocket universes with different values for the
constants of nature, is claimed to be well understood, for example
in case of eternal inflation
\cite{Vilenkin:1983xp,Linde4,Linde:1986fc}, but the problem is
that the volume of these thermalized regions with any given value
of the constants is infinite. Therefore, to compare them, one has
to introduce some cutoff and the results tend to be highly
sensitive to the choice of cutoff procedure
\cite{Linde:1993xx,Linde:1994gy,Linde:1995uf}. In a recent paper a
different method is proposed to find this distribution
\cite{Garriga:2005av}.

It should be stressed that this approach to the cosmological
constant problem is especially used within string theory, where
one has stumbled upon a wide variety of possible vacuum states,
rather than a unique one
\cite{PolchinskiBousso,Susskind1,Susskind2,Kachru:2003aw,Denef:2004ze,FreivogelSusskind,Douglas:2003um,Ashok:2003gk}.
By taking different combinations of extra-dimensional geometries,
brane configurations, and gauge field fluxes, a wide variety of
states can be constructed, with different local values of physical
constants, such as the cosmological constant. These are the 3-form
RR and NS fluxes that can be distributed over the 3-cycles of the
Calabi Yau manifold. The number of independent fluxes therefore is
related to the number of 3-cycles in the 6-dimensional  Calabi Yau
space, and can be several hundred. In addition, the moduli are
also numerous and also in the hundreds, leading to a total number
of degrees of freedom in a Calabi Yau compactification of order
1,000 or more. The number of metastable vacua for a given Calabi
Yau compactification therefore could be $10^{1000}$, and the
spacing between the energy levels $10^{-1000}M_{P}^4$, of which
some $10^{500}$ would have a vacuum energy that is anthropically
allowed. The states with (nearly) vanishing vacuum energy tend to
be those where one begins with a supersymmetric state with a
negative vacuum energy, to which supersymmetry breaking adds just
the right amount of positive vacuum energy. This picture is often
referred to as the ``Landscape''. The spectrum of $\rho_\Lambda$
could be very dense in this `discretuum' of vacua, but nearby
values of $\rho_\Lambda$ could correspond to very different values
os string parameters. The prior distribution would then no longer
be flat, and it is unclear how it should be calculated.

A review of failed attempts to apply anthropic reasoning to models
with varying cosmological constant can be found in
\cite{Garriga:2000cv}. See \cite{SmolinAnthrop} for a recent
critique. Another serious criticism was given in
\cite{Aguirre:2001zx}, where it is argued that very different
universes than our own could also lead to a small cosmological
constant, long-lived stars, planets and chemistry based life, for
example a cold big bang scenario. An analysis of how to make an
anthropic prediction is made in \cite{Aguirre:2005cj}.

A not very technical and almost foundational introduction to the
anthropic principle is given by \cite{Linde6}.

\subsubsection{Discrete Anthropic Principle}

It might be worthwhile to make a distinction between a continuous
anthropic principle and a discrete version. Imagine we have a
theory at our hands that describes an ensemble of universes
(different possible vacuum solutions) with different discrete
values for the fine structure constant:
\begin{equation}
\frac{1}{\alpha}=n+\mathcal{O}\left(\frac{1}{n}\right)
\end{equation}
such that the terms $1/n$ are calculable. An anthropic argument
could then be used to explain why we are in the universe with $n
=137$. Such a version of the anthropic principle might be easier
to accept than one where all digits are supposed to be
anthropically determined. Note that we are already very familiar
with such use of an anthropic principle: In a finite universe,
there is a finite number of planets and we live on one of the
(very few?) inhabitable ones. Unfortunately, we have no theory at
our hands to determine the fine structure constant this way, let
alone the cosmological constant.

\subsection{Summary}

This very much discussed approach offers a new line of thought,
but so far, unfortunately, predictions for different constants of
Nature, like the cosmological constant and the fine-structure
constant, are not interrelated. We try to look for a more
satisfying approach.

\section{Conclusions}

In this paper we categorized the different approaches to the
cosmological constant problem. The many different ways in which it
can be phrased often blurs the road to a possible solution and the
wide variety of approaches makes it difficult to distinguish real
progress.

So far we can only conclude that in fact none of the approaches
described above is a real outstanding candidate for a solution of
the `old' cosmological constant problem. Most effort nowadays is
in finding a physical mechanism that drives the Universe's
acceleration, but as we have seen these approaches, be it by
modifying general relativity in the far infrared, or by studying
higher dimensional braneworlds, generally do not convincingly
attack the old and most basic problem.

Since even the sometimes very drastic modifications advocated in
the proposals we discussed do not lead to a satisfactory answer,
this seems to imply that the ultimate theory of quantum gravity
might very well be based on very different grounds than imagined
so far. The only way out could be the discovery of a symmetry that
forbids a cosmological constant term to appear.

\ack Throughout this work I have benefited a lot from many
valuable discussions with my supervisor Gerard 't Hooft.

\section*{References}
\bibliographystyle{utcaps}
\bibliography{Categorizingbib}

\end{document}